# Single-domain chemical, thermochemical and thermal remanences in a basaltic rock

Ulrike Draeger[*], Michel Prévot, Thierry Poidras and Janna Riisager[**]

*Laboratoire de Tectonophysique, CNRS-UM2, Université de Montpellier 2,
34095 Montpellier Cedex 5, France.*

*Corresponding author: Michel Prévot (prevot@dstu.univ-montp2.fr)*

*present address: Riedener Weg 30, 82319 Starnberg, Germany
** now at: Geological Museum, University of Copenhagen, Oster Voldgade 5-7, DK-1350 Copenhagen K, Denmark.

Abbreviated title: Chemical and thermal remanences

November 3, 2005 17:08

**Abstract**

Tiny basaltic samples containing finely grained titanomagnetite with Curie temperature less than 100°C were heated in air in weak field (25 to 100µT) at temperatures between 400 and 560°C for times as long as 32 hours. Oxyexsolution of titanomagnetite resulted in the crystallization of interacting single domain particles with Curie point close to 540°C and the concomitant development of one of two types chemical remanence, depending upon thermal treatment: isothermal chemical remanence (CRM) or thermochemical remanence (TCRM), the latter acquired under the combined effects of chemical change and temperature decrease. CRM and TCRM acquired under various conditions were subjected to Thellier-type experiments. All these treatments were carried out using a vibrating sample thermomagnetometer allowing the continuous recording of magnetization and a very precise temperature control. All CRM-TRM and TCRM-TRM plots were found to be linear over almost the entire TRM blocking temperature range, whether pTRM checks are positive or not. An apparent strength of the acquisition field of CRM or TCRM could thus be obtained and divided by the actually applied field to obtain a ratio R which is representative of the CRM/TRM or TCRM/TRM ratios over most of the unblocking/blocking temperature spectrum. For CRM, it is found that R is less than one and increases rapidly with acquisition temperature ($0.36\pm0.07$ at 400°C, $0.66\pm0.02$ at 450°C, and $0.90\pm0.02$ at 500°C), in qualitative agreement with expressions derived from the theory of non-interacting single domain grains. Thus, very large underestimate of geomagnetic field paleostrength can occur when a natural CRM is not recognized as such and is believed to be a natural TRM. Paleointensity data obtained from geological material prone to the development of secondary minerals, as for example baked contacts and volcanic glasses, have therefore to be considered with caution. In an attempt to mimic deuteric oxyexsolution in cooling magma, TCRM was imparted from 560 to 400°C at a cooling rate of 0.1°C/mn. In contrast to the CRM case, R is found to slightly exceed one ($1.11\pm 0.05$), which is in agreement which theoretical considerations.

## 1. Introduction

It is a common feature in basaltic rocks that primary high-titanium titanomagnetite which crystallizes at high temperature subsequently undergoes some oxyexsolution during magma cooling. The highly magnetic mineral formed during this subsolidus chemical reaction is a very finely grained titanium-poor titanomagnetite. This is a magnetic carrier particularly appreciated in paleomagnetism because of the high stability of its remanences versus field, temperature and time. Moreover, this mineral is generally magnetically and chemically stable when heated at moderate temperatures. For all these reasons, volcanic rocks carrying oxyexsolved titanomagnetite are commonly preferred for studying the intensity of the Earth's magnetic field over geological time.

Paleointensity methods using volcanic rocks assume that NRM is a TRM. However, this fundamental assumption cannot be correct if oxyexsolution continues down to temperatures less than the Curie temperature of titanium poor titanomagnetite, that is commonly found to be around 550°C. An extensive magnetic study of magnetic oxides in two slowly cooling Hawaiian lava lakes led Grommé *et al.* (1969) to the conclusion that titanomagnetite oxyexsolution in nature occurs between 850 and 500°C, and even possibly down to 300°C. Typically, most of the blocking temperatures of oxyexsolved titanomagnetite are above 500°C. Therefore, there is a distinct possibility that the primary remanence acquired by some volcanic rocks is a thermochemical remanent magnetization (TCRM) rather than a TRM (Geissman and Van der Voo, 1980). In the present paper, we define a TCRM as a magnetization acquired under the combined influence of chemical change and temperature decrease. In contrast, we will use the expression "chemical remanent magnetization" (CRM) for referring to a magnetization acquired as a result of chemical changes occurring at any constant temperature.

The objective of the work reported in this paper is to compare some characteristics of TCRM, CRM and TRM imparted to basaltic rock samples in the laboratory. Previous works in that field are restricted to the CRM case. The pioneer experimental studies of Haigh (1958) and Kobayashi (1959) showed that the intensity of CRM acquired by magnetite while it forms at 300°C is much weaker than that of TRM. Later on, Stacey and Banerjee (1974) demonstrated theoretically the general validity of that conclusion for non-interacting single domain particles. They showed that the CRM/TRM ratio depends on the ratio of the blocking temperatures of CRM and TRM and reaches the maximum value of one when these temperatures are equal. Most of subsequent experimental data obtained either with synthetic minerals (Pucher, 1969; Ik Gie and Biquand, 1988; Stokking and Tauxe, 1990) or rocks (Hoye and Evans, 1975; Nguyen and Pechersky, 1985) confirmed that low-field CRM magnetite acquired in the 300-500°C range is generally largely inferior to TRM. However Pucher (1969) and Kellog et al. (1970) reported some cases in which CRM is equal to TRM. Obviously, the fact that CRM and TRM acquired in the same field can have largely different intensities may be the source of large error in paleointensity determination when a natural CRM is not recognized as such. Unfortunately, CRM and TRM cannot be easily distinguished from each other using usual paleomagnetic techniques. According to Kobayayshi (1959) and Pucher (1969) CRM and TRM generally have similar stability with respect to both AF and thermal demagnetization. Nevertheless McClelland (1996) suggested from theoretical considerations that CRM and TRM carried by non-interacting SD grains can be distinguished from each other on the basis of Thellier paleointensity experiments.

Our samples were selected from a rock containing very fine titanomagnetite particles, with the hope that the magnetic carriers exsolved during CRM or TCRM acquisition will be within the single domain range. This turned out to be the case, which allows an evaluation of our experimental data in light of SD theories. In this paper we will use the Néel theory (1949) of single domain grains. Although some aspects of this theory can be challenged, as we will discuss in the present paper, this is to date the only theoretical approach allowing to deal in a quantitative manner with most aspects of remanent magnetism in rocks, in particular CRM and TCRM intensities. In order to evaluate the reliability of paleointensity determinations, our study was focussed on the comparison of intensity and thermal unblocking/blocking of thermal (TRM) chemical (CRM) and thermochemical (TCRM) remanences. The main difficulty encountered in such studies is that, in contrast to deuteric oxyexsolution, oxyexsolution under laboratory conditions does not allow reaching a complete equilibrium of the newly formed phases with their oxidizing atmosphere. These exsolved phases are unstable during the subsequent heating needed for imparting a TRM. To try to evaluate the magnetic effects of the chemical changes occurring after CRM or TCRM acquisition, the TRM acquisition in the present work was carried out using the Thellier paleointensity method.

## 2. Experimental procedure and equipment

*2.1. Sample selection: paleomagnetic and initial rock magnetic experiments*

Several hand samples were taken from a very thin (about 15-20 cm thick) Quaternary basaltic dyke cutting a Jurassic limestone cliff close to the town of Carlencas (Southern France) along D908 (43.683°N, 3.228°E). Two hand samples (CAR4 and CAR5) were used for the present study. Seven small cores (12mm in diameter) were drilled out of hand sample CAR4, and 22 out of hand sample CAR5. Each drill core was then cut into up to seven 7mm-long specimens, labeled from "a" to "g" were "a" is the specimen close to the surface. Altogether about 120 specimens of about 2g in weight, 12mm in diameter and 7mm in length were labeled and used for our experiments. For example, specimen CAR51d is from hand sample CAR5, core number 1, and is the fourth specimen from the surface. In addition two one-inch diameter cores were drilled out of CAR5 in order to perform Thellier paleointensity experiments.

The NRM of the samples, measured using a JR5 spinner magnetometer, was progressively demagnetized up to 200 mT using a laboratory made alternating field demagnetizer, and the median alternating destructive field (MDF) was determined. Low field (0.1mT) thermomagnetic curves were obtained from some 40 specimens using either a Bartington susceptibility meter equipped with a furnace in which samples could be heated either in vacuum or in air or a Kappabridge in which specimens were heated in helium. The average Curie temperature of each magnetic phase was determined from the inflection point on the k-T curve (Prévot *et al.*, 1983). In complement, several samples were studied at the paleomagnetic laboratory in Munich to measure hysteresis cycle at room temperature and thermal variation of saturation magnetization Js.

40 fresh looking specimens were finally selected for CRM and TCRM acquisition experiments on the basis of the following criteria: low Curie temperatures (between 40 and 80 °C), MDF between 30 and 50 mT, and position in the hand sample as remote as possible from the surface (no "a" specimens were used).

*2.2. Acquisition of chemical remanences*

CRM and TCRM acquisition and measurement were carried out in air using a vibrating sample thermomagnetometer (for short, VTM in the following) manufactured by Orion (Yaroslavia, Russia) and modified by one of us (T. P.) in respect to temperature control and magnetic signal sampling. The magnetic sensor is composed of two identical detection coils connected in opposition in order to cancel ambient field fluctuations with zero or first order gradients. This magnetometer is surrounded by two concentric shield screens made of Mumetal in order to try to cancel the ambient field at the center of the magnetometer. Before the experiments reported here, the static residual field was checked and found to be less than 50nT in the entire sample displacement space. The sample is alternatively moved from one detection coil to the other with a frequency of 13.7 Hz and an amplitude of 25.4mm generating an electromotive force which is filtered and preamplified before data acquisition by the computer. The magnetization is measured continuously, each registered data being sampled over 14s. The non-inductive heater is made of two oppositely wound resistor wires driven by a Pulse Width Modulated power supply. The base frequency of this PWM power supply is 3740 Hz. The control of the temperature is done by an Eurotherm 808 controller and the temperature is measured with a thermocouple S placed in a little hole in the sample. Our software makes it possible to define 3 successive thermal stages, each with their proper heating or cooling rate and applied dc field, through the serial port of the computer connected to an Eurotherm controller. A solenoid located between the detection coils and the furnace enables us to apply a continuous inducing field up to 1.2mT upon the sample during the experiment.

In our experiments, the thermal cycle for CRM acquisition was carried out in air and consisted of three stages:

- stage 1: rapid heating (10°C/mn) from $T_0$ (room temperature) to Tr (temperature of CRM acquisition); the same field was applied during stages 1 and 2 in order not to impart a high temperature IRM to the sample at the beginning of stage 2;
- stage 2: constant temperature, typically maintained (within 1°C) for 32 hours in a 100μT dc field;
- stage 3: rapid cooling (10°C/mn) in zero field from Tr to $T_0$.

Approximately 30 CRM acquisition experiments were carried out, most of them at Tr=400°C, and a few at 450 or 500°C. A few experiments were carried out in steady fields of 25 or 50μT. Two experiments were made under a field of 100μT which was regularly turned off and on for a few minutes. These latter experiments were used to determine the ratio of remanent to induced magnetization, and for estimating short-term magnetic viscosity. These samples were not used to estimate the CRM/TRM ratio.

In order to monitor changes in magnetic mineralogy during CRM experiments, a distinct set of 13 fresh tiny specimens was heated to 400°C, each for a different duration distributed between 1 and 32 hours, using the same VTM equipment and procedure at Montpellier laboratory. Subsequently, hysteresis curves at 400°C were measured on these specimens at the Saint Maur geomagnetic laboratory using a laboratory-made translation inductometer (Bina and Daly, 1994) including a furnace and a field coil providing fields up to 0.2T. The maximum field we used for these hysteresis curves was up to 190mT.

The thermal cycle for TCRM acquisition, also carried out in air, consisted of the three following stages:
- stage 1: rapid heating (10°C/mn) in zero field from $T_0$ (room temperature) to 560°C, a temperature which exceeds the Curie point of oxidized samples by some 20°C (§ 4.1);
- stage 2: slow cooling (0.1°C/mn) in a 100μT constant field down to 400°C;
- stage 3: rapid cooling (10°C/mn) in the same field from Tr to $T_0$.

Three tiny specimens were subjected to this treatment. For four others the field was alternatively turned on and off every 47 or 68s.

*2.3. Thellier experiments on chemical remanences*

The Thellier experiments (Thellier and Thellier, 1959) were carried out in air using our modified VTM. TRM and CRM (or TCRM) were directed along the length of the tiny specimen, all with the same sense. We used the Coe variant (Coe, 1967) of the Thellier method, and we carried out pTRM checks (Thellier and Thellier, 1959) at every temperature steps. For each sample, the applied field was the same as for CRM or TCRM acquisition. For each Thellier run, all remanence measurements were carried out at the same temperature. For CRM, this measurement temperature was the same as the temperature of CRM acquisition (400, 450 or 500°C). For TCRM, the measurement temperature was the lowest temperature of the interval of TCRM acquisition (400°C). Thus, no magnetic phase with Curie temperature less than 400 to 500°C can possibly be involved in Thellier experiments.

Due to software constraints, temperature steps had to be constant for each Thellier run. The first experiments carried out with CRM involved temperature steps of 20°C. For better resolution of the pseudo-paleointensity data, these steps were subsequently reduced to 10°C, and finally to 4°C for the latest experiments. In parallel, the first temperature step was increased from 410°C to 490°C, since no significant CRM decrease or pTRM acquisition had been observed within the 400-490°C interval. For the Thellier experiments on TCRM, we used temperature steps of 10°C from 410°C to 560°C (or 580°C). The rate of increase or decrease of temperature employed was 4°C/mn. When the desired temperature was reached it was held for 5mn. The measured temperature drift never exceeded 1°C.

*2.4. Final rock magnetic experiments*

The objectives of these experiments were to characterize the initial magnetic phase, the phases formed during CRM acquisition, and subsequent changes possibly induced by repetitive heatings at higher temperatures. The following experiments were made either at the Montpellier laboratory, at the Saint Maur laboratory, at CEREGE (CNRS and Aix-Marseille University) or at the Institute for Rock Magnetism (IRM) at Minneapolis University, MN:

1) low and high field thermomagnetic curves (Montpellier and IRM). Low field experiments carried out at both places used a Kappabridge (field intensity 0.4mT). High field experiments used the MicroVSM (maximum field intensity 0.5T) at the IRM.
2) room temperature hysteresis measurements (Saint Maur, IRM and CEREGE). Bulk hysteresis data were obtained either at Saint Maur using the laboratory-made translation inductometer described above or at the IRM using a MicroVSM (maximum field intensity 1T). Differential hysteresis measurements such as first order reversal curves (FORC) were obtained at CEREGE from measurements by a Micromag 3900 VSM.
3) thermal variation of hysteresis characteristics (Saint Maur and IRM). Same equipment as described in item 2.
4) AF demagnetization of pTRM(520, 515°C) (Montpellier and Saint Maur). This pTRM was imparted using the Orion VTM, and then demagnetized at room temperature using similar laboratory-made alternating field demagnetizers.

## 3. Experimental results

*3.1. Starting material*

The low-field thermomagnetic curves of unheated material (e.g. CAR517g in Fig.1) exhibit Curie temperatures varying from 35 to 80 °C. This indicates titanomagnetite with high titanium content (x between 0.7 and 0.8). Hysteresis parameters plot in the central area of the PSD domain for Ti-rich titanomagnetite (Fig.2, samples CAR517f, CAR517g). Microscopy study of polished sections revealed a bimodal grain-size distribution of titanomagnetite with modes around 2 μm and 20 μm. No ilmenite was observed, neither as discrete crystals nor as exsolution within titanomagnetite crystals.

The two one-inch cores (CAR51B and CAR52C) were studied by the Thellier stepwise heating method. They do not exhibit any change in NRM direction during heating throughout the investigated temperature range. PTRM checks are positive up to 175 °C and the NRM-TRM plot is linear up to 190°C (Fig. 3). The NRM fraction is 0.7 and the quality factor q (Coe et al., 1978) is close to 25. Such long linear NRM-TRM plots obtained from high titanium titanomagnetite are usually typical of very fine grains (Prévot et al., 1983).

To better characterize the micromagnetic structure of magnetic carriers, both in fresh and in heated materials, two kinds of differential hysteresis analyses were carried out: FORC and differential IRM measurements. FORC diagrams (Pike et al., 1999; Roberts et al., 2000) represent the density distribution of irreversible differential magnetic moments in function of two fields mathematically defined (Roberts et al., 2000), and usually labeled Hc and Hu. For a set of randomly oriented identical uniaxial single-domain particles with axial coercive force hc, the average Hc value obtained from FORC should approach 0.48 hc, the theoretical value of bulk coercive force, and Hu should be a local interaction field (Néel, 1954). This interpretation is reasonable agreement with experimental FORC diagrams measured from well-characterized SD materials (Pike et al., 1999; Carvallo et al., 2004) or computed from various SD models (Pike et al., 1999; Carvallo et al., 2003; Muxworthy et al., 2004).

The interpretation of FORC diagrams provided by MD and PSD grains is more uncertain. As first proposed by Néel (1942), hysteresis due to wall movement can be formally treated as resulting from the combination of various types of elementary hysteresis cycles, as those represented in a Preisach-Néel diagram. However, this is a formal interpretation of magnetization curves to consider with caution. In particular, FORC diagrams obtained for MD grains reveal a more complex pattern than those which can be computed from simple MD models (Pike et al., 2001b). Therefore, the physical interpretation of the FORC coordinates remains unclear. Empirically, judging from the diagrams obtained by Roberts et al. (2000), rocks containing PSD and MD particles provide somewhat similar FORC diagrams characterized by a density of differential moments which is maximum at the origin, and a similar distribution of both Hu's and Hc's.

Magnetic after-effect (magnetic viscosity), which can affect all grain sizes from MD to small SD, is the source of some artifacts in FORC diagrams, as a secondary peak in moment distribution near the origin (Pike et al., 2001a) or, in our opinion, anomalies in moment distribution in the neighboring of the diagonal of the inferior quadrant of FORC diagrams (as in the middle diagram of Fig. 4). We think that the latter anomalies are probably due to the fact that this diagonal is the locus where the remagnetizing field Hb changes sign during the acquisition of FORC data.

The FORC diagrams of fresh specimen CAR521d (upper two diagrams on Fig. 4) look intermediate between those of SD and PSD/MD grains. On one hand, the peak of moment distribution is located at the origin, on the other hand Hc distribution reaches values exceeding 150mT while interaction fields are smaller by more than one order of magnitude. The weakness of interaction fields is independently corroborated by the small values of differential IRM (Petrovsky and al., 1993; Wehland et al., 2005). Differential IRM is the difference between the IRM measured in back field H after saturation in the opposite sense and the IRM value expected from Wolfarth's (1958) relations between IRM acquisition and SIRM d.c. demagnetization for non-interacting SD grains. Non zero differential remanence is attributed to interaction fields. Positive (negative) values indicate a net-magnetizing (net-demagnetizing) effect of magnetic interactions. For specimen CAR521d, differential IRM, particularly scattered in low back fields, remains close to zero for any applied field H (Fig. 5). Altogether, bulk hysteresis and differential hysteresis data suggest that the magnetic configuration of the fresh material is mostly PSD.

*3.2 Chemical remanence acquisition*

The VTM provides a continuously record of total magnetization J changes at temperature as experiment proceeds. When exposed to a non-zero field, the measured magnetization is the sum of a remanent and an induced component. The latter, calculated from field on/field off experiments, was found to represent only 2-3% of the total magnetization for acquisition temperatures up to 450°C, and 5-9% for higher temperatures (acquisition of CRM(500°C) and TCRM). Let us note that the percentages given above are approximately overestimated by a factor of two since the applied field is off during half of the duration of the CRM or TCRM acquisition process. The viscosity of remanence (viscosity index; Prévot, 1981) measured over a few minutes, was found to be lower than 1%. Thus J measurements provide a very good approximation of Jcrm or Jtcrm, even though a small high-temperature VRM component is inevitably included in these measurements.

During the first stage of the thermal cycle of CRM acquisition (rapid heating to Tr under a constant field), J is found to decrease rapidly to be reduced to a few per cent of the room temperature magnetization at temperatures around 150-200°C (Fig. 6a). Beyond 200°C a small progressive increase is observed. This increase in J reflects essentially an increase in susceptibility related to the onset of chemical changes around 200°C: field on/field off experiments indicate that at any temperature between 200 and 400°C, J is essentially an induced magnetization. At the onset of the temperature dwell at 400°C, only 25% of J is a remanent magnetization. Thus the CRM acquired during heating from 200 to 400°C is extremely small compared to the final CRM (Fig. 6 a and b; note that the J scale differs by one order of magnitude between these two diagrams). CRM develops rapidly as soon as temperature is stabilized at 400°C. Field on/field off experiments indicate that only a few minutes after the onset of this temperature dwell, CRM reaches already 75% of J.

Typically, CRM growth follows approximately a log t law up to the end of the constant temperature stage (32hrs) (Fig. 6b and 7a). This simple behavior was also reported by Gapeev et al. (1991) for CRM resulting from low-temperature oxidation of nearly single-domain magnetite. In fact, more detailed calculations show that in our case CRM acquisition passes through a maximum one hour after the beginning of the constant temperature stage (Fig. 7b). An approximately log t acquisition law corresponds the simplest behavior which can be derived from the Néel (1949) relaxation theory of single-domain particles. However, for about 1/3 of our CRM acquisition experiments, a marked S-shape was observed in J as a function of log t. These samples were not studied any longer.

Additional experiments were carried out on a separate set of 8 specimens that were heated at 400°C, each of them for a distinct duration between 2 and 32 hrs. When normalized to the same inducing field, CRM intensity is found to reach that of NRM in less than two hours.

The remanence of samples meant for TCRM experiments decays rapidly during heating to 560°C in zero field (Fig. 8a). In this diagram, the change in sign around 250°C is an instrumental artifact due to a progressive drifting of the baseline at the high sensitivity required for these measurements. During the subsequent very slow cooling in a 100µT field, TCRM acquisition starts only around 540°C (Fig. 8b). The interpretation of the shape of the TCRM acquisition curve is delicate because of concomitant changes in temperature and time. A tentative calculation of the TCRM blocking curve will be discussed in §4.3.

*3.3. Thellier experiments on chemical remanences*

Examples of CRM-TRM diagrams are given in Fig. 9 for CRMs acquired at 400°C or 500°C, and TCRM-TRM diagrams are shown in Fig. 10. The data for all samples are listed in Table 1 following the canon of presentation of Thellier paleointensity data (Coe et al., 1978). An additional column gives R, the ratio of the apparent paleointensity to the actual acquisition field. For this calculation, it was assumed that CRM is proportional to applied field in the investigated range (25 to 100µT).

All plots are almost perfectly linear over large CRM fractions f (from 0.61 to 0.95). f is essentially dependent on the magnitude of the low-temperature CRM decrease (from 400°C to, say, 510°C) in which (CRM,TRM) couples do not fall in line with higher temperature data. As we will see below (§5), the unblocking/blocking processes are ineffectual in that low-temperature interval. The changes in CRM and TRM which are depicted on the diagrams of Fig. 9 and 10 in that interval result essentially from mineralogical transformations. Moreover, those changes are dependent on the mode of heating from 400 to 490°C (Fig. 16 below).

The quality factor q is largely controlled by the number of heating steps. Considering the data obtained using 4°C temperature intervals, q is about 30 for TCRM while it varies from 12 to 64 for CRM. Linearity of the diagrams derived from Thellier experiments is sometimes considered as an argument in favor of the absence of magnetic alteration during experiments (e.g. Stokking and Tauxe, 1990). Obviously, this is not necessarily the case here. If we consider the 4 best documented data set for CRM, obtained with 4°C heating steps, only two samples yield positive pTRM checks (like sample CAR56d in Fig. 9). Checks are fairly good for the second (sample CAR55e, Fig. 9) but they are clearly negative for the last one (sample CAR51d, Fig. 9): all checks carried out in the TRM unblocking interval indicate a significant increase in pTRM at each temperature step. The three samples carrying a TCRM (Fig. 10) behave rather similarly to the last CRM-carrying sample. The origin and the possible effect on R of the changes occurring during Thellier experiments, below and within the temperature interval in which plots are linear, will be examined in section 5.

In practice, the R ratio listed in Table 1 is a measure of the CRM/TRM (or TCRM/TRM) strength ratio over some 90% of the TRM blocking range (Fig. 9 and 10). For CRM's, R is approximately constant at a given temperature of acquisition Tr. For the best documented case (Tr = 400°C) R = 0.36±0.07 for six determinations (in this paper, ± refers to standard deviation). Within the investigated range of Tr (400-500°C) R is always less than one and increases almost linearly with acquisition temperature (Fig. 11). In contrast, R calculated from TCRM's is slightly but systematically larger than one.

## 4. Process of chemical remanence acquisition

*4.1. Mineral process*

In Fig. 1 we can compare the thermomagnetic curves, both in low and high-field, for a fresh sample (CAR517g) and a sample previously heated in air at 400°C for 32 hrs (CAR518e). As usual, low field curves are much more discriminative for distinguishing coexisting magnetic phases. After CRM acquisition, the magnetic phase presents in the fresh material has almost entirely disappeared. Instead, three magnetic phases with high Curie temperature are observed. Their respective average Curie temperatures determined by the method of Prévot et al. (1983) are 522 ± 11°C for high T phase 1, 450 ± 10°C for high T phase 2, and 390 ± 30°C for high T phase 3. Other low field experiments on different samples showed that phases 2 and 3 are not always distinct from each other and exhibit somewhat variable Tc, although they remain in the range 350-500°C. Such phases may be either titanomaghemite or titanomagnetite. When samples having acquired a CRM are repetitively heated above 400°C, an intermediate phase with Curie point around 440°C is often observed (Fig. 12). Although this phase progressively diminishes while phase 1 grows, it does not unmix at high temperature as expected for titanomaghemite. Thus, such intermediate Tc phases correspond probably to titanomagnetite with x between 0.2 and 0.4. Phase 1, that is favored by high temperature treatments, seems to correspond to the dominant near-stoichiometric magnetite found in intergrowths resulting from inversion of titanomaghemite (Tucker and O'Reilly, 1980; Özdemir, 1987). Combining the data obtained from different samples subjected to various high temperature treatments, we estimate the final Curie temperature to be approximately 540°C. This is compatible with a stoichiometric titanomagnetite with x = 0.06 and $Js_0$ = 421kA/m (Kawai, 1956).

Progressive magnetic mineral changes during CRM acquisition were monitored using a set of 13 fresh specimens heated to 400°C, each for a different duration between 1 and 32 hours, using the VTM equipment at Montpellier laboratory (§2.2). The data obtained from the hysteresis measurements at 400°C, subsequently carried out at Saint Maur laboratory, are listed in Table 2 and shown in Fig. 13. Although each data point corresponds to a distinct sample, the magnetic homogeneity of our sample set (see §2.1 for selection criteria) is such that coherent trends versus heating time are documented for all hysteresis data ( Js, Jrs, Hcr and Jrs/Js) with the only exception of Hcr/Hc which remains stationary. In spite of the small irregularities due to the combination of data from distinct samples, three main stages can be readily distinguished during the CRM acquisition process:
- stage 1 (from 1 to 2 hours): large increase in both induced and remanent saturation magnetizations without noticeable changes in coercivity, Jrs/Js or Hcr/Hc (see Fig. 2 for comparison with fresh material);
- stage 2 (from 2 to 4 hours): steep increase in Jrs/Js ratio from 0.30 to 0.40-0.45 accompanied by an increase in saturation magnetizations and coercive forces;

- stage 3 (beyond 4 hrs): saturation magnetizations continue to increase while both Jrs/Js ratio and coercivity remain almost constant.

In addition, other experiments carried out on another set of specimens (§3.2) showed that isothermal heating to 400°C increase the median destructive AF of CRM. Comparable to that of NRM for t = 2hrs (36mT), the MDF of CRM reaches 50-60mT for heating times between 20 and 30 hours.

Stage 1 may correspond to a monophase oxidation producing metastable titanomaghemite (Goss, 1988) with progressively increasing oxidation degree and Curie point. The crucial step in the mineralogical process described above is the decrease in magnetic grain size which occurs during stage 2. This reduction is expected in cases of exsolution, whether it occurs by spinodal decomposition or nucleation and grow (Fig. 14). The brevity of this stage attests to the simultaneity of mineralogical reactions affecting magnetic grains within the whole specimen. After this stage, no significant change in magnetic grain size is documented although some slight increase in volume might be suspected for the largest heating times. The range of the Jrs/Js ratio during stage 3 is 0.40-0.45. These values are only slightly smaller than the theoretically value of 0.5 expected for SD grains with shape (and therefore uniaxial) anisotropy. Let us remember that, in order to avoid any contribution from the magnetic phases not involved in the CRM acquisition process, this ratio was measured at 400°C, not at $T_0$. For reasons which remain unclear, Jrs/Js generally tends to diminish with temperature. This decrease is about 20% for naturally oxyexsolved titanomagnetite heated to 400°C (Dunlop, 1987). The Jrs/Js ratio found for those samples at that temperature seems therefore indicative of a truly SD magnetic configuration. Beyond 10 hrs or so the coercivity remains constant while saturation magnetizations continue to grow. As we will see below, CRM seems essentially carried by phase 1. Considering the large saturation magnetization of this phase, the microcoercive force Hc of these SD particles is due to shape anisotropy. Assuming that magnetization reversal is coherent, we have then :

$$Hc = Nd\,Is \qquad (1)$$

where Is is the spontaneous magnetization and Nd the demagnetizing factor of the particle. Thus, if the Js increase was due to an increase in spontaneous magnetization, then Hc should grow with Js. The absence of correlation between Hc and Js suggests that during this final stage of CRM acquisition the increase in Js is mainly due to volume growth of the magnetized material (within the space limited by the ilmenite frame) while the Is value, and therefore the chemical composition, of the final magnetic phase remains unchanged. This suggestion is confirmed by low-field thermomagnetic experiments such as those shown in Fig. 12 that show that repetitive heating to high temperatures leads to a further development of the phase 1 mineral without any significant increase in Curie temperature. Thus, both hysteresis changes during CRM acquisition and thermomagnetic observations suggest that the oxyexsolution process results from nucleation and growth rather than from spinodal decomposition (Fig. 14).

The mineral process at work during TCRM acquisition from 560 to 400°C is certainly the same as during isothermal CRM acquisition since this temperature range is well within the domain of oxyexsolution of titanium-rich titanomagnetite (O'Reilly, 1984). There is, however, a distinct characteristic of TCRM acquisition when compared to CRM acquisition: all things being equal, the reaction rate diminishes drastically with time in the first case, while it is constant in the second case. For an activation energy of 240kJmol$^{-1}$, which is a typical value for cation diffusion process in minerals (Putnis, 1992), the reaction rate is divided by 4,000 as T decreases from 560 to 400°C. From 560 to 500°C, the reaction rate is already reduced by an order of magnitude. This has an important consequence: during the process of TCRM acquisition, most of the chemical changes are expected to occur within the highest temperature range.

*4.2. Final size and shape of CRM carriers*

Fig. 2 demonstrates that the CRM carriers are much smaller in size than the original high-titanium titanomagnetite crystals. However, these measurements carried out at room temperature cannot be used to estimate the size of CRM carriers because of the presence in the heated sample of a remnant fraction of the original low Curie temperature phase (Fig.1a). As pointed out in the previous section, the hysteresis measurements carried out at 400°C during CRM acquisition suggest that the final magnetic product of the chemical changes is of SD size. This inference is clearly confirmed by the FORC diagram (lower diagram of Fig. 4) obtained at room temperature from sample CAR57b (previously heated for 32 hours at 400°C, then used for various measurements at temperatures between 400°C and 580°C). This diagram exhibits only a single phase of high coercivity displaying typical SD characteristics. The peak in coercivity distribution is about 53mT. Following Néel's interpretation of the hysteresis cycle of SD particles oriented at random (1954), the coercive force of individual particles measured along the magnetization axis is 110mT. A further important information from this diagram is the presence of large interaction fields, as expected within magnetite-ilmenite intergrowths (e.g., Davis and Evans, 1976). The magnitude of the interaction field defined by the FWHM on the FORC peak (Muxworthy et al., 2004) is 45mT. Moreover, the plot of the differential IRM versus the back applied field obtained from sample CAR57b shows a large negative peak (Fig. 5), which confirms the presence of interactions and suggests that they are dominantly negative.

A quantitative estimate of the size and axial coercivity of typical CRM carriers can be obtained from the SD theory by combining low-field thermoviscous blocking and AF unblocking experiments. This method, based on Néel (1949) theory of single-domain non-interacting particles, was used for thermoremanence (Dunlop and West, 1969) and viscous remanence (Biquand et al., 1971; Prévot, 1981). Bina and Prévot (1977) showed that hematite grain sizes calculated with this method are in reasonable agreement with direct size measurement from TEM observations. According to Dunlop and West (1969), the presence of magnetic interactions does not invalidate such calculations since we are dealing with low-field remanences. They argue that only the grains subjected to very small interaction fields could have their magnetic moments aligned by such applied fields. In practice, the method consists of solving simultaneously the blocking and unblocking equations.

In low field the blocking expression at some decreasing temperature $T>T_0$ is:

$$vH_{c0} = \frac{2kT\,(Q+\ln\tau_b)}{I_{s0}\,F(T)\,F'(T)} \qquad (2)$$

with $F(T) = \frac{I_s(T)}{I_{s0}}$ and $F'(T) = \frac{H_c(T)}{H_{c0}}$

The unblocking conditions under a large field H applied at room temperature $T_0$ are:

$$\frac{v(H_{c0}-H)^2}{H_{c0}} = \frac{2kT_0(Q+\ln\tau_u)}{I_{s0}} \qquad (3)$$

In these expressions k is Boltzman's constant, and v, Is and Hc are volume, spontaneous magnetization and microscopic coercive force of particle, respectively. Following Néel's (1949) notations, Q is a constant estimated to 22 for magnetite, $\tau_b$ is the blocking time (depending upon cooling rate) and $\tau_u$ the unblocking time (depending upon AF frequency).

The thermal unblocking of CRM(400°C) as measured at 400°C is generally maximum around 520°C. As we simply wish to get an approximate estimate of size and coercivity of CRM carriers rather than a complete description of the density distribution of grains in the whole (Hc, v) diagram, a single pTRM was imparted to sample CAR57b, between 515 and 520°C, before FORC experiments. This pTRM was then progressively AF demagnetized at room temperature. The blocking temperature to be introduced in expression (2) is the mean temperature of the pTRM acquisition interval. Two distinct AF experiments provided the same median alternating destructive field (140mT). This value of the unblocking field was introduced into expression (3). From the experimental blocking and unblocking procedures, we estimated that $\tau_b \cong 15s$ and $\tau_u \cong 6 \times 10^{-6}s$. F(T) and F'(T) between $T_0$ and T could not be precisely measured on our samples because of the presence of several magnetic phases (e. g. Fig.1 and 12). Thus, those ratios were taken from previous experiments on a thermally stable basalt sample containing deuteritic oxyexsolved (titano)magnetite with Curie temperature close to that of phase 1 (Dunlop et West, 1969; Dunlop, 1987). This is probably the main source of uncertainty in our calculation. Other experimental sources of imprecision are the thermal instability of our samples and the fact that the temperature interval chosen for pTRM acquisition may correspond to grains which are not perfectly representative of the CRM carriers as a whole.

Our calculations give $H_{c0}$ = 144mT and v = 2.3 x $10^{-3}\mu m^3$. This microcoercive force is of the same order of magnitude as the mean axial coercive force calculated from the FORC diagram. As mentioned above, phase 1 is strongly magnetic and the coercivity can be assumed to be due to shape anisotropy. If coherent reversal of magnetic moment is assumed, then equation 1 is valid. Given the dependence of Nd on the particle dimension ratio (see for example Nagata, 1961), the calculated microcoercive force requires a dimension-ratio close to two. Thus, considering the calculated volume mentioned above, the particle length would be about 0.2µm.

The dimension ratio obtained from equation (1) is possibly a minimum estimate. According to theoretical considerations, coherent rotation in magnetite probably occurs only in a restricted size range below 50nm (Dunlop and Özdemir, 1997). The mode of magnetization reversal of larger elongated particles, like the present CRM carriers, is believed to be incoherent (Moon and Merril, 1988; Enkin and Williams, 1994), which in turn should reduce particle coercivity. Thus a larger dimension ratio of the CRM-carrying particles would be needed to account for the calculated axial coercivity. The reduction in coercivity depends both on the grain size and the mode of incoherent reversal (buckling, curling or fanning; see for example Dunlop and Özdemir, 1997). Theoretical considerations suggest that prolate spheroid may present only two modes of incoherent reversal, curling and buckling, the latter one being restricted to extremely large aspect-ratios (Aharoni, 2000) which are not realistic for natural crystals. Thus, the most probable non-coherent reversal mode of elongated magnetite particles is curling (Newell and Merril, 1999). Taking into account the volume of the CRM-carrying particles, curling would reduce their coercivity by a factor of 3 as compared to that given by equation 1 (Luborsky, 1961). To account for the observed CRM coercivity, the dimension ratio should be then increased to 10-20.

It must be noted however that we have no experimental evidence in favor of the occurrence of incoherent reversal of magnetization in the final, largest-size CRM carriers. Whichever is the mineralogical process giving birth to the magnetite-ilmenite intergrowth (Fig. 14), a growth in the size of the magnetic phase has to occur during CRM acquisition experiments. According to the theoretical inferences presented above, the reversal mode of the CRM-carrying particles should evolve with time. Coherent rotation should predominate at the onset of our CRM acquisition experiments, the particle size being then just above the SP/SD threshold. Then curling should progressively becomes predominant since the final grain size is clearly above the SD/PSD threshold for equidimensional particles. Thus, the coercivity of the CRM-carriers should markedly decrease with time, especially during the latest stage of CRM acquisition. However, the observed trend in coercivity described in §4.1 does not corroborate that expectation: coercive forces remain constant during that stage (Fig.

13). It would seem therefore that our experimental data are at variance with some theoretical inferences based on numerical micromagnetics, namely the change from coherent to incoherent reversal mode of magnetization when the grain size of elongated single-domain magnetite particles grows from the SP to the PSD thresholds.

The present state-of-the-art does not allow investigating the origin of this discrepancy. On one side, there are the shortcomings (like neglecting the effects of defects) and all the unsolved difficulties of micromagnetics computations which makes "any comparison with experiments of the computational results for perfect particles… meaningless and misleading" (Aharoni, 2000). On the other side, the quantitative constraints of the exsolution-and-growth mineralogical model of oxyexsolution need to be computed, in particular the dynamics of the propagation of oxyexsolution in the sample and individual titanomagnetite crystals, and the evolution of the shape of magnetite intergrowths during the transformation process.

*4.3. Blocking of chemical and thermochemical remanences*

All previous theoretical studies on CRM acquisition dealt with grain-growth CRM (Stacey and Banerjee, 1974; Shcherbakov et al., 1996; McClelland, 1996; Stokking and Tauxe, 1990). The present case is more complex: both mineralogical models of exsolution (Fig. 14) indicate that both chemical composition and volume of magnetic particles do change with time, which is confirmed by the hysteresis measurements reported above (§4.1). Moreover the increase in volume is strictly limited by the ilmenite lamellae framework.

As pointed out by Néel (1949), regardless of the physical quantity acting (field, temperature, time etc…), magnetic blocking can be described as resulting from a single process, the increase with time of the relaxation time τ given by:

$$\frac{1}{\tau} = C \exp\left(-v H_c I_s / 2kT\right) \quad (4)$$

where the coefficient $C \cong 10^{10} s^{-1}$ (Aharoni, 1992) estimates the frequency of "attempts" to overcome the potential barrier that prevents reversal of the moment.

In the present case, magnetization blocking is governed by the way titanium progressively migrates from the spinel towards and into the hexagonal corundum structure. Whichever way exsolution develops (Fig. 14), two magnetic blocking processes are at work simultaneously: (i) a progressive increase of the volume of titanium-poor titanomagnetite with a given chemical composition, and (ii) a progressive increase in Fe/Ti ratio within a given volume. The increase in Fe/Ti ratio raises spontaneous magnetization, which in turn increases the microscopic coercive force in proportion. Thus, chemical change and volume increase work together towards an increase of the relaxation time.

Let us examine the following two limiting cases of CRM blocking: blocking by chemical changes with an unchanged volume of magnetic mineral (Is-blocking model) or volume-growth blocking with an unchanged chemical composition (v-blocking model). In the Is-blocking model, $v = 2 \times 10^{-3} \mu m^3$ is constant, and blocking at 400°C occurs then spontaneous magnetization reaches the value $I_s(b) = 23 kA/m$. This is compatible with a titanomagnetite having an ulvospinel molecular fraction x close to 0.26, a Curie temperature near 415°C, and $I_{s0} = 306 kA/m$ (Kawai, 1956). In the v-blocking model, $I_s = 257 kA/m$ is constant, and the blocking volume at 400°C is $v(b) = 0.2 \times 10^{-3} \mu m^3$. For a dimension ratio equal to 2, this volume corresponds to a length of about 90nm. Considering the development with time of titanium concentration (Fig. 14), the spinodal decomposition model tends to favor the first magnetic blocking process, while the second magnetic blocking process is more compatible with the nucleation and growth model.

In the case of TCRM blocking, all variable terms in equation 4 vary simultaneously, each of them acting as to increase relaxation time. Thus, it is difficult to separate the role of each process (chemical change, volume increase and temperature decrease) with regard to magnetic blocking. However, some constraints can be derived from thermodynamic and magnetic considerations. As pointed out above, the chemical reaction rate dependence on T is so strong that most of changes clearly occur within the highest temperature range of the TCRM experiments. Moreover, the *per descendum* temperature trend during blocking requires the blocked magnetic volumes to correspond, at any temperature, to the highest Curie temperatures then present in the sample. The blocking temperature spectrum (at 400°C) of the TCRM acquired by sample CAR56b has been calculated from the TCRM acquisition curve (Fig. 8b) after correction by the thermal variation of Js (measured on the adjacent sample CAR57b previously heated for 32 hours at 400°C). The TCRM blocking curve so obtained is shown on Fig. 15. This curve is somewhat approximate because of the various sources of imprecision in the Js(T) correction. In particular, the progressive slight decrease in TCRM that seems to occurs from 450 to 400°C is believed to be an artifact. However, the interval of TCRM blocking is clearly established: more than 90% of remanence is blocked between 540 and 500°C. This indicates that TCRM is carried by phase 1 mineral. The magnetic measurements represented on Fig. 8b show that this phase starts forming rapidly, within 3 hours from the onset of the high temperature treatment.

## 5. Changes during Thellier experiments

Under laboratory conditions intended to simulate natural deuteric oxidation, titanomagnetite is not able to reach equilibrium with the oxidizing environment within a reasonable time scale, unless temperature exceeds 1,000°C (Tucker and O'Reilly, 1980). It is thus expected that the magnetic carriers of CRM or TCRM will continue to undergo chemical transformations during the Thellier experiments, since these latter experiments are carried out in the same oxidizing atmosphere but at higher temperatures than the CRM and TCRM acquisition experiments. Indeed, changes were observed to occur both below and within the unblocking/blocking range in which the CRM-TRM or TCRM-TRM diagrams are linear.

Sample alteration at temperatures below the temperature interval providing a linear plot is indirectly attested by the steeper slopes provided by the CRM-TRM or TCRM-TRM data obtained below that interval (Fig. 9 and 10). A few additional experiments have been carried out to investigate the changes occurring during Thellier experiments for samples carrying a CRM. Upon progressive heating from 400°C using small constant ΔT intervals, no significant variation in CRM intensity is generally observed until CRM starts to unblock around 490-500°C (Fig. 16a). This suggests that there is no noticeable contribution from intermediate Curie temperature phases to the CRM. Consequently, for the subsequent Thellier experiments on CRM, the first temperature step chosen was 490°C. At the same time, beyond this temperature, each step was reduced to 4°C in order to increase the number of data in the unblocking interval of CRM. Fig. 16b shows an example of the unexpected behavior then observed in the range 490-500°C. As the heating temperature is stepwise increased, a steep increase in CRM and a somewhat symmetric decrease in TRM occur. From 490 to 502°C, the TRM is successively reduced to 66%, 58%, 28% and 12%, that in spite of the increase of its blocking temperature range. To avoid any interference with thermal blocking/unblocking processes, another experiment was carried out, the sample being repeatedly heated at the same temperature (490°C). As seen in Fig. 17a, the TRM(490°, 400°) is initially different from zero and vanishes progressively as the number n of heating increases. In contrast, the CRM(490°C, Tc) tends to increases with n. As can be seen in Fig. 17b, the decrease of the acquired pTRM varies in proportion to the increase of the CRM left.

A possible first-order explanation of these observations is an increase in the volume of individual magnetic particles during Thellier experiments. This growth might be a mere continuation of the similar trend documented during the latest stage of CRM acquisition (§4.1). According to equation 4, the expression of the blocking temperature is:

$$T_b = \frac{v\, Hc\, Is}{2\, k\, \ln(C\, \tau_{th})} \quad (4\text{bis})$$

where $\tau_{th}$ is the blocking time, which depends on cooling rate. This equation shows that a very small volume increase results in a noticeable shift of $T_b$ towards higher values. At the temperatures under consideration, a 1% increase in v would increase $T_b$ by 8°C. From a theoretical standpoint, an increase in volume of individual SD particles results in a similar increase in remanence intensity, regardless of the kind of remanence considered (CRM, TCRM or TRM). Since CRM and TRM have the same direction and polarity in our experiments, the CRM increase may be explained by the overstepping of the 490°C boundary by particles formerly acting as TRM carriers. This shift should also decrease the intensity of the low temperature pTRM, providing that the source of blocking temperatures less than 490°C is depleted.

As seen above, the occurrence of chemical changes within the TRM blocking range (typically 510-540°C for CRM and 470-540°C for TCRM) is indicated -both in some of the CRM-carrying samples and in the three TCRM-carrying samples- by a systematic increase in pTRM capacity (Fig. 9 and 10). This trend may also be interpreted as due to an increase in particle volume. At each temperature step of Thellier experiments, chemical changes take place essentially during the first heating cycle (field off). If the volume, and thus the magnetic moment of each magnetic grain increases, so do both the pre-existing and newly acquired remanences in proportion, as long as the single domain structure is maintained. Thus, even though sample magnetic mineralogy does evolve from one temperature step to the next, it is conceivable that the TCRM/CRM ratio might not be noticeably modified. Actually, the chemical changes occurring during Thellier experiments do not seem to alter significantly the CRM/TRM ratio. For Tr = 400°C, this ratio is found to be 0.36±0.05 for the two samples for which no noticeable pTRM increase is observed during Thellier experiments. It is equal to 0.36±0.08 for the four other.

## 6. Comparing chemical and thermal remanences: from theory to practice

*6.1. Theoretical considerations*

According to Néel (1949), provided the blocking condition expressed by equation 4 is met, the blocked average magnetic moment of a set of identical non-interacting SD particles with their easy magnetization axis parallel to the applied field H, when measured at any temperature $T_0$ equal or less than $T_b$, is :

$$\overline{m_0} = v\, I_{s0}\, th\frac{v\, I_s\, H}{k\, T} \quad (6)$$

For a randomly oriented assembly integrated from 0 to ±π/2, and considering that the argument of the hyperbolic tangent is less than one in our case, the blocked average moment at $T_0$ is approximately given by:

$$M_0 = \frac{1}{3}\, v\, I_{s0}\, \frac{v\, I_s\, H}{k\, T} \quad (7)$$

Applying the blocking equation, we can replace $T = T_b$ by $\frac{v\, H_c\, I_s}{2k\, \ln(C\, \tau_{(b)})}$ where $\tau_{(b)}$ is the blocking time of the remanent magnetization under consideration. Also, since the coercive force is due to shape anisotropy in the present case, and provided magnetization reversals are coherent, Hc is equal to Nd Is. Then the remanence intensity is expressed by:

$$M_{R(0)} = \frac{2 H v I_{s0}}{3 N_d} \frac{\ln(C\tau_{(b)})}{I_s} \qquad (8)$$

For CRM, the expression may be written:

$$M_{CRM(0)} = \frac{2 H v I_{s0}}{3 N_d} \frac{\ln(C\tau_{(crm)})}{I_{s(crm)}} \qquad (9)$$

where, as usual (Néel, 1949; Aharoni, 1992), the blocking time $\tau_{(crm)}$ is taken to be of the same order as the time taken for the measurement (14s in the present case). In equation 9 all the quantities are initially associated with the magnetic carrier at the blocking instant. At this moment, v and Is are largely different for the two blocking models considered above (§5), and so is the CRM intensity. Subsequently however, since the mineralogical process is still going on and is evolving towards the same final mineral assemblage, CRM intensity converges towards the same value for the volume-blocking and the Is-blocking models.

If a TRM is now imparted to this final assemblage, and assuming that no changes occur during TRM acquisition, we have similarly:

$$M_{TRM(0)} = \frac{2 H v I_{s0}}{3 N_d} \frac{\ln(C\tau_{(trm)})}{I_{s(trm)}} \qquad (10)$$

where $I_{s(trm)}$ is the spontaneous magnetization of the final mineral produced during CRM experiments, measured at $T_{trm}$, the temperature of TRM blocking. $\tau_{(trm)}$, the blocking time for thermoremanence, is the time needed for $\tau$ to increase sufficiently for a superparamagnetic grain population initially close to the SP/SD threshold to become fully blocked (Néel, 1949; Dodson and McClelland-Brown, 1980). Thus $\tau_{(trm)}$ is inversely proportional to cooling rate. This is the origin of the dependence of TRM intensity upon cooling rate, first experimentally verified by Papusoi (1972).

It follows from equations 9 and 10 that:

$$\frac{M_{CRM}}{M_{TRM}} = \frac{I_{s(Ttrm)}}{I_{s(Tcrm)}} \frac{\ln(C\tau_{(crm)})}{\ln(C\tau_{(trm)})} \qquad (11)$$

This equation is equivalent to the expressions obtained by Stacey and Banerjee (1974), Stokking and Tauxe (1990) and Shcherbakov et al. (1996) who all neglected the logarithmic term.

For the TCRM case, we have similarly:

$$\frac{M_{TCRM}}{M_{TRM}} = \frac{I_{s(Ttrm)}}{I_{s(Ttcrm)}} \frac{\ln(C\tau_{(tcrm)})}{\ln(C\tau_{(trm)})} \qquad (12)$$

where $\tau_{(tcrm)}$ is the blocking time for TCRM.

It must be remained that equations 8 to 10 are correct only if the mode magnetization reversal of particles is coherent. If not, the magnetic moment calculated from these expressions should be too low. However, as discussed in §4.2, some of ours experimental evidences are difficult to conciliate with the occurrence of non-coherent reversal in the CRM-carrying particles. In any case, the occurrence of incoherent magnetization reversals would not affect the validity of equations (11) and (12) since they express the ratio of two magnetic moments.

*6.2. Comparing experimental data and theoretical predictions*

Let us first compare experimental versus theoretical remanence ratios. The experimental TCRM/TRM average is 1.11±0.05. According to the VTM measurements carried out on sample CAR56b (Fig. 15), the median blocking temperatures of TCRM and TRM are very close to each other. Considering the uncertainties in the method of calculation of the TCRM blocking curve (§4.3) and the fact that repetitive heating does not seem to modify the Curie temperature of phase 1 mineral (Fig. 12), the TCRM and TRM blocking temperatures may be assumed to be equal. Regarding the blocking time $\tau$, Néel (1949) proposed to define it as the time needed for the relaxation time to be multiplied by $e$. Thus, in our case $\tau$(tcrm) is inversely proportional to both the cooling rate and the rates of Is increase and volume growth. Since the rates of the physical and chemical changes of grains are unknown, a maximum estimate of $\tau$(tcrm) is used, based on the cooling rate alone (0.1°C/mn). The TRM cooling rate is 4°C/mn. Using those values of Tb's and $\tau$'s, one obtains a theoretical TCRM/TRM ratio equal to 1.1, which is in agreement with experimental findings. However, this perfect coincidence is probably fortuitous : any small difference between Tb(tcrm) or Tb(trm) would result in largely different theoretical TCRM/TRM ratios.

For samples carrying a CRM, the experimental CRM/TRM ratio is found to increase with temperature Tr (0.36±0.07 at 400°C, 0.66±0.02 at 450°C, and 0.90±0.02 at 500°C), which is in agreement with the trend expected from equation 11. These experimental findings seem to be of general validity. Nguyen and Pechersky (1985) reported a CRM/TRM ratio equal to 0.47 for a CRM acquired by apparently single domain magnetite resulting from oxyexsolution of large-sized basaltic titanomagnetite heated in air for 26 hours at 400°C. In the absence of Thellier experiments, Nguyen and Pechersky (1985) calculated the CRM/TRM ratio from the values of initial CRM and final TRM. According to our Thellier experiments, this method provides CRM/TRM ratios which are larger (by 20-25%) than those obtained from the linear part of CRM-TRM diagrams (Fig. 9). Thus their experimental data and ours are in good agreement. We can note in passing that there is no reason for the size of the exsolved phases to be dependent on the size of the host titanomagnetite. Rather, it should be generally expected that exsolution size and spacing are mainly dependent on the temperature of oxyexsolution because of the strong dependence of the rate of ion diffusion on temperature.

The theoretical CRM/TRM values were calculated using a median TRM blocking temperature of 525±5°C, as typically found from our Thellier experiments. As can be seen on Fig. 11, theoretical ratios are systematically and significantly lower than experimental determinations: 0.12±0.04 at 400°C, 0.17±0.05 at 450°C, and 0.39±0.11 at 500°C.

For CRM, another inference from the theory of magnetism of non-interacting single domain is not experimentally verified. The very same equation 11 that qualitatively accounts for the observed trend in CRM/TRM ratio as the blocking temperature of CRM is modified, also implies that the Thellier experiments provide concave up CRM-TRM plots. The concavity expected from theory is large enough to be easily seen in Fig. 9. For a CRM acquired at 400°C, the slope of the curve in the TRM blocking range should be equal to 0.26 near the lower limit (typically 510°C) and while it should be 0.02 near the upper limit (typically 540°C). Actually, all our samples exhibit a nice linear plot over almost the entire TRM blocking range, whether pTRM checks are positive or not.

*6.3. Possible effects of magnetic interactions*

The differential hysteresis experiments carried out at room temperature, especially the FORC diagram, demonstrate the presence of large magnetic interactions in between the single-domain CRM-carrying particles. This conclusion certainly holds for TCRM-carrying particles as well. It is well established that magnetic interactions can largely reduce the intensity of low-field remanences such as TRM or ARM (Dunlop and West, 1969), and a similar effect has to be expected for CRM. In the present context, which we wish to know is whether magnetic interactions modify the *ratio* of CRM to TRM. Shcherbakov et al. (1996) addressed this topic by carrying out a comparative Monte Carlo simulation of CRM and TRM acquisition by SD interacting uniaxial magnetite grains. For a crystallization temperature of 300°C, they found that CRM/TRM ratios less than one were typical of non-interacting grains with volume concentration less than 1%. For concentrations larger than 1-3%, CRM/TRM ratios equal to or larger than one were computed. In our basaltic samples, the almost-magnetite formed by oxyexsolution (our phase 1) is clustered within the former Ti-rich titanomagnetite crystals where its concentration is very large. Considering that these initial Fe-Ti-O crystals contained some 70-80% of ulvospinel before oxidation (§3.1), the local volumetric concentration of magnetite should lie in the range 20-30%. According to the theoretical results of Shcherbakov et al.'s (1996), a CRM/TRM ratio close to one or more has therefore to be expected for our samples, with consequently no dependence with the temperature of CRM acquisition. These predictions are at odds with our experimental findings and those of Nguyen and Pechersky (1985) as well.

However, the simple qualitative considerations developed below suggest that magnetic interactions can explain the two major discrepancies noted above between experimental data and predictions from the non-interacting single-domain theory of magnetism: (i) experimental CRM/TRM ratios systematically larger than theoretical ones; and (ii) experimental CRM-TRM diagrams linear (Fig. 9) rather than concave up.

Whichever is the mineralogical scenario of oxyexsolution considered (Fig. 14), weak magnetic interactions are expected at the instant when the direction of the CRM-carrying SD moment gets blocked. In case of spinodal decomposition, the remanence-carrying particles have then a very weak spontaneous magnetization (§4.3). In case of nucleation and growth, the blocked particles are strongly magnetic, but the intragranular distance is generally much larger than the grain diameter. Subsequently, spontaneous magnetization or particle size increases so that interaction field growths up, reaching 45mT after completion of thermal treatments (§4.2). These interactions are about three times weaker than particle axial coercivity and cannot significantly reduce the intensity of the previously blocked CRM. But they should reduce the intensity of the TRM initiated later in this highly interactive magnetic configuration.

The linearity of experimental CRM-TRM diagrams can also result from magnetic interaction. The increase of these fields as oxyexsolution proceeds should result in a random re-distribution of the intrinsic blocking temperatures of grains. This suggestion is in agreement with Néel's (1949) theory of magnetism of SD particles which implies that the energy barrier E of a particle depends on external fields. E increases when the external field He is parallel to the magnetic moment of particle and decreases when they are antiparallel, the amplitude of the change in E increasing with the magnitude of He. This results in qualitatively similar shifts of $T_b$. When subjected to strong interaction fields of different sign and magnitude, a set of identical single domain grains with an initially unique intrinsic $T_b$ yields a broad distribution of $T_b$'s around the initial value. According to our experiments, interaction fields appear large enough for redistributing $T_b$'s all over their intrinsic (interaction-free) range of variation.

Regarding TCRM, the agreement between experimental and theoretical remanence ratios suggests that large interaction fields are already established before TCRM blocking. This suggestion implies that oxyexsolution develops largely before blocking, that is to say during the three-hour cooling from 560°C to the upper limit of the TCRM blocking range, close to 540° C (Fig. 15). This view is supported by the fact that at a temperature as low as 400°C, the breaking down of titanomaghemite into a magnetite/ilmenite assemblage does not take more than 3 hours, while theoretical considerations suggest that the reaction rate in the 540-560°C range is larger than that at 400°C by 2 to 3 orders of magnitude (§4.1). It is thus reasonable to assume that most of mineralogical changes do occur before TCRM blocking. The behavior of TCRM should then tends towards that of a simple TRM, in agreement with the fact that the TCRM/TRM ratio is found to be equal to one (once the cooling rate difference is taken into account) and the TCRM versus TRM plot is linear.

## 7. Conclusions

The present study suggests that isothermal oxyexsolution process in basaltic titanomagnetite at moderate temperatures (500-400°C) results from nucleation and growth rather than from spinodal decomposition. Tiny Ti-poor titanomagnetite domains of nanometric size form rapidly along the boundaries of nucleating ilmenite lamellae. The magnetization of these domains gets blocked in the absence of noticeable magnetic interactions, due to the large distance between them. The orientation of magnetic moments remains constant while the size of the single domain Ti-poor titanomagnetite develops within the ilmenite lamellae framework and interaction field strongly increase. The final product is an assemblage of strongly coercitive (110-140mT) single-domain particles with strong magnetic interactions between them (around 45mT). As a result of these local interaction fields, the intrinsic blocking/unblocking temperatures of individual particles are modified in a random fashion. In our experiments, this effect results in linear CRM/TRM curves, in contradiction with the concave up shape expected from the theory of magnetism for non-interacting single-domain grains. The alteration of Tb distribution during oxyexsolution help to explain Walderhaug's (1992) observation that the genuine direction of the primary remanence of rocks heated in a weak field for several hours at moderate temperatures (400 to 525°C) can no longer be retrieved after CRM acquisition. This overlap of the blocking temperatures can result from the redistribution of the blocking temperatures of the primary remanence under the influence of stronger and stronger interaction fields. In nature, long-term heating at moderate temperatures are expected to occur typically near igneous contacts. It is often observed that primary and secondary magnetizations cannot be separated from each other in the baked zone, as pointed out by Camps et al. (1995) for the Steens Basalt sequence. This suggests that such magnetic overprints are low-temperature chemical remanences rather than partial thermoremanences.

In agreement with previous experimental works (Nguyen and Pechersky, 1985; Stokking and Tauxe, 1990), our experiments show that the intensity of the isothermal CRM acquired at a moderate temperature Tr by single domain particles is largely inferior to that of the total TRM acquired above this temperature. The CRM/TRM ratio is shown here to be strongly dependent on temperature Tr, with values from less than 0.4 at 400°C to 0.9 at 500°C. These findings are in qualitative agreement with the theory of non-interacting single domain particles. However, the experimental ratios are significantly larger than theoretical values, which we tentatively attribute to the effect of magnetic interactions.

Knowing that CRM can provide a linear CRM-TRM plot when studied by the Thellier method invites particular caution in paleointensity studies on geological materials which may have experienced late mineral crystallization at moderate or low temperatures, as for example baked contacts or volcanic glasses. Because they generally provide nice linear NRM-TRM diagrams, baked contacts are often considered as an ideal material for paleointensity studies. However, considering how moderate is the temperature reached during basement heating by a cooling lava or dyke (Audusson & Levi, 1988; Camps *et al.*, 1995; Kristjansson, 1985) and the long duration of this process, there are good reasons to have some doubts about the actual nature of NRM. Volcanic glasses constitute another material of debatable reliability for paleointensity studies. Prone to rapid alteration at water temperature (Jakobsson, 1978; Moore, 1966), they are also thermodynamically unstable and can undergo post-eruption devitrification (e.g. Marshall, 1961). Submarine basalt glasses (SBG) generally provide linear NRM-TRM plots (Pick and Tauxe, 1993) and are thus particularly appreciated for paleointensity studies. However, it is puzzling that the remanence of SBG is carried by some almost stoichiometric magnetite (Pick and Tauxe, 1994) instead of high-titanium titanomagnetite, the pre-deuteric magnetic phase systematically present in fresh submarine pillow basalts (Schaeffer and Schwarz, 1970; Prévot et al., 1979) and continental tachylites as well (Smith and Prévot, 1977). It is well established (e.g. O'Reilly, 1984) that Ti-poor titanomagnetite can crystallize at sea-floor temperature and thus carry a CRM rather than a TRM. In contrast, basaltic high-titanium titanomagnetite have crystallization temperature largely exceeding their Curie temperature.

Heller et al. (2002) showed that the Earth's mean dipole moment calculated from the paleointensity data obtained from SBG over the last 20Ma is unexpectedly 40% less than the moment calculated from the data provided by continental volcanic rocks, although the main magnetic carriers are typically Ti-poor titanomagnetite in both cases. Thus, these authors suggested that a significant fraction of SBG provides too low paleointensity determination due to a low-temperature crystallization of magnetic carrier. The results described in the present paper give a solid experimental support to their suggestion.

Most of paleointensity data from continental volcanic rocks are obtained from basaltic flows containing titanomagnetite oxyexsolved as a result of deuteric alteration during magma cooling. The TCRM experiments reported in the present paper are the first attempt to mimic this natural phenomenon in order to evaluate its possible effects on paleointensity determination.  Our experimental data suggest that titanomagnetite oxyexsolution concomitant with magnetization blocking results in TCRM/TRM ratios larger than one by some 10%. Theoretical considerations suggest that this overestimate might be simply explained by the difference in cooling rates during TCRM and TRM acquisition processes in the laboratory. In nature, it is probable that oxyexsolution develops mostly at temperatures higher than the TRM blocking range. In this case, no bias due to magnetic interaction should affect paleointensity data. However, the difference in cooling rate between the natural and laboratory blocking processes has to be taken into account according to expression 12.

## 8. Acknowledgements

We thank our colleagues at the Saint Maur geomagnetic laboratory, especially Mansour Bina, and at the Institute for Rock Magnetism (IRM) in Minneapolis, MN, for helping U.D. during her stays in these laboratories. The choice of the rock used here was based on low-field thermomagnetic studies previously carried out by Elisabeth Schnepp, who kindly provided some of her material to us. We acknowledge Pierre Rochette and Fabienne Vadeboin at CEREGE (University Aix-Marseille and CNRS) who kindly carried out FORC and differential IRM experiments and computation, and Liliane Faynot for her help in drawing figures. Ken Hoffman kindly revised an early version of this paper and made both linguistic and scientific suggestions for improvement. Reviews by David Dunlop, Ron Merrill and an anonymous referee led to substantial improvements in the manuscript. This research was supported by a postdoctoral fellowship from the ROKMAG European Network (number ERBCHRXCT93-0315) funded by the Commission of the European Communities.

# References

Aharoni, A., 1992. Relaxation processes in small particles, *in Studies of magnetic properties of fine particles and their relevance to materials sciences*, pp. 3-11, ed. Dormann, J. L. & Fiorani, D., Elsevier.

Aharoni, A., 2000. *Introduction to the theory of ferromagnetism,* 2nd edn, Oxford University Press, Oxford, United Kingdom, 319 pp.

Audunsson, H. & Levi, S., 1988. Basement heating by a cooling lava: paleomagnetic constraints, *J. geophys. Res.*, **93**, 3480-3496.

Bina, M. & Daly, L., 1994. Mineralogical change and self-reversed magnetizations in pyrrhotite resulting from partial oxidation; geophysical implications, *Phys. Earth planet. Inter.*, **85**, 83-99.

Bina, M. & Prévot, M., 1977. Hematite grains: size and coercive force from AF demagnetization at high temperatures, *Phys. Earth planet. Inter.*, **13**, 272-275.

Biquand, D., Prévot, M. & Dunlop, D.J., 1971. Une aimantation rémanente visqueuse très résistante envers les champs alternatifs dans une roche sédimentaire contenant de fines particules d'hématite, *J. Phys.*, **32**, supplement to 2-3, Colloque C1, 1043-1044.

Camps, P., Prévot M. & Coe, R., 1995. Revisiting the initial sites of geomagnetic impulses during the Steens Mountain polarity reversal, *J. geophys. Res.*, **123**, 484-506.

Carvallo, C., Muxworthy, A. R., Dunlop, D.J. &Williams, W., 2003. Micromagnetic modeling of first-order reversal curve (FORC) diagrams for single-domain and pseudo-single-domain magnetite, *Earth planet. Sci. Lett.*, **213**, 375-390.

Carvallo, C., Özdemir, Ö. & Dunlop, D.J., 2004. First-order reversal curve (FORC) diagrams of elongated single-domain grains at high and low temperatures, *J. geophys. Res.*, **109**, 10.1029/2003JB002539.

Coe, R., 1967. Paleo-intensities of the Earth's magnetic field determined from Tertiary and Quaternary rocks, *J. geophys. Res.*, **72**, 3247-3262.

Coe, R., Grommé, S. & Mankinen, E.A., 1978. Geomagnetic paleointensities from radiocarbon-dated lava flows on Hawaii and the question of the Pacific nondipole low, *J. geophys. Res.*, **83**, 1740-1756.

Davis, P.M. & Evans, M.E., 1976. Interacting single-domain properties of magnetite intergrowths, *J. geophys. Res.*, **81**, 989-994.

Day, R., Fuller M. & Schmidt, V.A., 1977. Hysteresis properties of titanomagnetites: grain size and composition dependence, *Phys. Earth Planet. Inter.*, **65**, 62-77.

Dodson, M.H. & McClelland-Brown, E., 1980.Magnetic blocking temperatures of single-domain grains during slow cooling, *J. geophys. Res.*, **85**, 2625-2637.

Dunlop, D.J., 1987. Temperature dependence of hysteresis in 0.04-0.22µm magnetite and implications for domain structure, *Phys. Earth planet. Inter.*, **46**, 100-119.

Dunlop, D.J., 2002. Theory and application of the Day plot (Mrs/Ms versus Hcr/Hc) 1. Theoretical curves and tests using titanomagnetite data, *J. geophys. Res.*, **107** (B3), 10.1029/2001JB000486.

Dunlop, D.J. & Özdemir, Ö., 1997. *Rock magnetism: fundamentals and frontiers*, Cambridge University Press, Cambridge, United Kingdom, 573 pp.

Dunlop, D.J. & West G. F., 1969. An experimental evaluation of single domain theories, *Rev. Geophysics*, **7**, 709-757.

Enkin, R.J. & Williams, W., 1994. Three-dimensional micromagnetic analysis of stability in fine magnetic grains, *J. geophys. Res.*, **99**, 611-618.

Gapeev, A.K., Gribov, S.K., Dunlop, D.J., Ozdemir, O. & Shcherbakov, V.P., 1991. A direct comparison of the properties of CRM and VRM in the low-temperature oxidation of magnetite, *Geophys. J. Int.*, **105**, 407-418.

Geissman, J.W. & van der Voo, R., 1980. Thermochemical remanent magnetization in Jurassic silicic volcanics from Nevada, *Earth planet. Sci. Lett.*, **48**, 385-396.

Goss, C.J., 1988. Saturation magnetization, coercivity and lattice parameter changes in the system $Fe_3O_4$-$\gamma Fe_2O_3$, and their relationship to structure, *Phys. Chem. Minerals*, **16**, 164-171.

Grommé, C.S., Wright, T.L. & Peck, D.L., 1969. Magnetic properties and oxidation of iron-titanium oxide minerals in Alae and Makaopuhi lakes, Hawaii, *J. geophys. Res.*, **74**, 5277-5293.

Haigh, G., 1958. The process of magnetization by chemical change, *Phil. Mag.*, 3, 267-286.

Heller, R., Merrill, R.T. & McFadden, P.L., 2002. The variation of intensity of earth's magnetic field with time, *Phys. Earth planet. Inter.*, **131**, 237-249.

Hoye, G.S. & Evans M.E., 1975. Remanent magnetizations in oxidized olivine, *Geophys. J. R. Astr. Soc.*, **41**, 139-151.

Ik Gie, T. & Biquand D., 1988. Etude expérimentale de l'aimantation rémanente chimique acquise au cours des transformations réciproques hématite-magnétite, *J. Geomag. Geoelectr.*, **40**, 177-206.

Jakobsson, S., 1978. Environmental factors controlling the palagonitisation of the Surtsey tephra, Iceland, *Bull. Geol. Soc. Den.*, **27**, 91-105.

Kawai, N., 1956. Subsolidus phase relation in titanomagnetite and its significance in rock-magnetism, *Proceed. Intern. Geol. Congress, Mexico,* **XI-A**, 103-120.

Kellog, K., Larson, E.E. & Watson, D.E., 1970. Thermochemical remanent magnetization and thermal remanent magnetization: comparison in a basalt, *Science,* **170**, 628-630.

Kobayashi, K., 1959. Chemical remanent magnetization of ferromagnetic minerals and its application to rock magnetism, *J. Geomag. Geoelectr.*, **10**, 99-117.

Kristjansson, L., 1985. Magnetic and thermal effects of dike intrusions in Iceland, *J. geophys. Res.*, **90**, 10,129-10,135.

Luborsky, F.E., 1961. High coercive materials. Development of elongated particles magnets, *J. Appl. Phys.,* **32**, 171S-183S.

Marshall, R.R., 1961. Devitrification of natural glasses, *Geol. Soc. Am. Bull.*, **72**, 1493-1520.

McClelland, E., 1996. Theory of CRM acquired by grain growth, and its implications for TRM discrimination and paleointensity determination in igneous rocks, *Geophys. J. Int..*, **126**, 271-280.

Moon, T. & Merril R. T., 1988. Single-domain theory of remanent magnetism, *J. geophys. Res.*, **93**, 9202-9210.

Moore, J.G., 1966. Rate of palagonitization of submarine basalt adjacent to Hawaii, *U.S. Geol. Surv. Prof. Pap.,* **550-D**, D163-D171.

Muxworthy, A., Heslop, D. & Williams, W., 2004. Influence of magnetostatic interactions on first-order-reversal-curve (FORC) diagrams: a micromagnetic approach, *Geophys. J. Int..*, **158**, 888-897.

Newell, A.J. & Merrill, R.T., 1999. Single-domain critical size for coercivity, *J. geophys. Res.*, **104**, 617-628.

Nagata, T., 1961. *Rock magnetism*, rev. edn, pp.70-71, Maruzen, Tokyo.

Néel, L., 1942. Théorie des lois d'aimantation de Lord Rayleigh. 1ère partie: les déplacements d'une paroi isolée, *Cahiers Phys.* **12**, 1-20.

Néel, L., 1949. Théorie du traînage magnétique des ferromagnétiques en grains fins avec applications aux terres cuites. *Ann. Géophys.*, **5**, 99-136.

Néel, L., 1954. Remarques sur la théorie des propriétés magnétiques des substances dures, *Appl. Sci. Res.,* **B-4**, 13-24.

Nguyen, T. K. T. & Pecherskiy, D.M., 1985. Criteria for remanent crystallization magnetization in magnetite containing rocks, *Izvestiya, Earth Phys. (English translation)*, **21**, 633-641.

O' Reilly, W., 1984. *Rock and mineral magnetism*, 1st edn, pp.22-26, Blackie & Son, Glasgow.

Özdemir, Ö, 1987. Inversion of titanomaghemite, *Phys. Earth planet. Inter.*, **46**, 184-196.

Papusoi, C., 1972. Effet de la vitesse de refroidissement sur l'intensité de l'aimantation thermorémanente d'un ensemble de grains monodomaines, *Ann. Sci. Univ. "Al I. Cuza", Phys.*, **18**, 31-47.

Petrovsky, E., Hedja, P., Zelinka, T., Kropacek, V. & Subrt, J., 1993. Experimental determination of magnetic interactions within a system of synthetic haematite particles, *Phys. Earth planet. Inter.*, **76**, 123-130.

Pick, T. & Tauxe, L., 1993. Holocene paleointensities: Thellier experiments on submarine basaltic glass from the East Pacific Rise, *J. geophys. Res.*, **98**, 17,949-17,964.

Pick, T. & Tauxe, L., 1994. Characteristics of magnetite in submarine basaltic glass, *Geophys. J. Int.*, **119**, 116-128.

Pike, C.R., Roberts, A.P. and Verosub, K.L., 1999. Characterizing interactions in fine magnetic particle systems using first order reversal curves, *J. Appl. Phys.*, **85**, 6660-6667.

Pike, C.R., Roberts, A.P. and Verosub, K.L., 2001a. First-order reversal curve diagrams and thermal relaxation effects in magnetic particles, *Geophys. J. Int.*, **145**, 721-730.

Pike, C.R., Roberts, A.P., Dekkers, M.J. and Verosub, K.L., 2001b. An investigation of multi-domain hysteresis mechanisms using FORC diagrams, *Phys. Earth planet. Inter.*, **126**, 11-25.

Prévot, M., 1981. Some aspects of magnetic viscosity in subaerial and submarine volcanic rocks. *Geophys. J. R. Astr. Soc.*, **66**, 169-192.

Prévot, M., Lecaille, A. & Hékinian, R., 1979. Magnetism of the Mid-Atlantic ridge crest near 37°N from FAMOUS and D.S.D.P. results : A review, in *Deep Drilling results in the Atlantic Ocean : Ocean crust*, pp. 210-229, ed. Talwani, M., Harrison, C.G. & Hayes, D.E., Maurice Ewing serie, 2, Amer. Geophys. Union.

Prévot, M., Mankinen, E.A., Grommé, C.S. & Lecaille, A., 1983. High paleointensity of the geomagnetic field from thermomagnetic studies on Rift Valley pillow basalts from the Mid-Atlantic Ridge, *J. geophys. Res.*, **88**, 2316-2326.

Pucher, R., 1969. Relative stability of chemical and thermal remanence in synthetic ferrites, *Earth planet. Sci. Lett.*, **6**, 107-111.

Putnis, A., 1992. *Introduction to mineral sciences*, 1st edn, pp. 275-331, Cambridge Univ. Press.

Roberts, A. P., Pike, C.R. & Verosub, K.L., 2000. First-order reversal curve diagrams: a new tool for characterizing the magnetic properties of natural samples, *J. geophys. Res.*, **105**, 28,461-28,475.

Schaeffer, R.M. & Schwarz, E.J., 1970. The mid-Atlantic ridge near 45°N. IX. Thermomagnetics of dredged samples of igneous rocks, *Can. J. Earth Sci.*, **7**, 268-273.

Shcherbakov, V. P. & Sycheva, N.K., 1996. Monte Carlo modelling of TRM and CRM acquisition and comparison of their properties in a ensemble of interacting grains, *Geophys. Res. Lett.*, **23**, 2827-2830.

Smith, B.M. & Prévot, M., 1977. Variation of the magnetic properties in a basaltic dyke with concentric cooling zones, *Phys. Earth Planet. Inter.*, **14**, 120-136.

Stacey, F.D. & Banerjee, S.K., 1974. *The physical principles of rock magnetism,*1st edn, pp.128-135, Elsevier.

Stokking, L.B. & Tauxe L., 1990. Properties of chemical remanence in synthetic hematite: testing theoretical predictions, *J. geophys. Res.*, **95**, 12,639-12,652.

Thellier, E. & Thellier, O., 1959. Sur l'intensité du champ magnétique terrestre dans le passé historique et géologique. *Ann. Géophys.,* **15**, 285-376.

Tucker, P. & O'Reilly, W.O., 1980. The laboratory simulation of deuteric oxidation of titanomagnetites: effect on magnetic properties and stability of thermoremanence, *Phys. Earth planet. Inter.*, **23**, 112-133.

Walderhaug, H., 1992. Directional properties of alteration CRM in basic igneous rocks, *Geophys. J. Inter.*, **111**, 335-347.

Wehland, F., Stancu, A., Rochette, P., Dekkers, M. & Appel, E, (2005). Experimental determination of magnetic interaction in natural and artificial pyrrhotite bearing samples, *Geophys. J. Int.*, in print.

Wohlfarth, E.P.,1958. Relations between different modes of acquisition of the remanent magnetization of ferromagnetic particles, *J. Appl. Phys.*, **29**, 595-596.

| | Laboratory remanence acquisition | | | Thellier experiments | | | | | | |
|---|---|---|---|---|---|---|---|---|---|---|
| Sample | Type | $T_r$, °C | Field, µT | $\Delta T$, °C | N | f | g | q | $H_{app} \pm \sigma(H_{app})$, µT | $R \pm \sigma(R)$ |
| CAR55e | CRM | 400 | 25 | 514 - 538 | 7 | 0.65 | 0.80 | 12 | 7.1 ± 0.3 | 0.28 ± 0.01 |
| CAR53d | CRM | 400 | 50 | 510 - 540 | 4 | 0.82 | 0.58 | 8 | 14.9 ± 0.9 | 0.30 ± 0.02 |
| CAR44a | CRM | 400 | 100 | 500 - 560 | 4 | 0.95 | 0.50 | 5 | 43.5 ± 3.9 | 0.44 ± 0.04 |
| CAR53e | CRM | 400 | 100 | 500 - 540 | 5 | 0.91 | 0.62 | 12 | 42.2 ± 2.0 | 0.42 ± 0.02 |
| CAR56d | CRM | 400 | 100 | 514 - 534 | 6 | 0.61 | 0.78 | 16 | 39.5 ± 1.2 | 0.40 ± 0.01 |
| CAR513d | CRM | 400 | 100 | 506 - 534 | 8 | 0.83 | 0.82 | 64 | 32.6 ± 0.3 | 0.33 ± 0.01 |
| CAR44b | CRM | 450 | 100 | 500 - 560 | 4 | 0.85 | 0.37 | 10 | 66.1 ± 2.0 | 0.66 ± 0.02 |
| CAR51d | CRM | 500 | 100 | 514 - 542 | 8 | 0.75 | 0.85 | 43 | 88.9 ± 1.3 | 0.89 ± 0.01 |
| CAR45b | TCRM | 560 - 400 | 100 | 470 - 540 | 8 | 0.86 | 0.81 | 27 | 107.5 ± 2.7 | 1.08 ± 0.03 |
| CAR56b | TCRM | 560 - 400 | 100 | 470 - 540 | 8 | 0.88 | 0.79 | 33 | 108.9 ± 2.3 | 1.09 ± 0.02 |
| CAR56c | TCRM | 560 - 400 | 100 | 490 - 540 | 6 | 0.77 | 0.73 | 31 | 116.5 ± 2.1 | 1.17 ± 0.05 |

**Table 1:** Results of Thellier experiments on CRM and TCRM (experimental procedure described in §2.3). $T_r$: temperature of CRM or TCRM acquisition; $\Delta T$: temperature interval used for apparent paleointensity determination; N: number of temperature steps used for determination; f, g, q: CRM (or TCRM) fraction used for apparent paleointensity determination, gap factor, and quality factor, respectively (Coe et al., 1978); $H_{app}$: apparent paleointensity value with associated error; R: ratio of the apparent paleointensity to the actual acquisition field.

| Sample | Heating time at 400 °C | $J_{rs}$, $10^{-3}$ Am²/kg | $J_s$, $10^{-3}$ Am²/kg | $J_{rs}/J_s$ | $H_c$, mT | $H_{cr}$, mT | $H_{cr}/H_c$ |
|---|---|---|---|---|---|---|---|
| CAR517e | 1 hr | 22 | 70 | 0.314 | 5.3 | 10 | 1.89 |
| CAR52d | 2 hrs 16 min | 37 | 121 | 0.306 | 5.0 | 8.8 | 1.76 |
| CAR58f | 3 hrs 19 min | 43 | 124 | 0.347 | 7.4 | 10.6 | 1.43 |
| CAR519e | 4 hrs | 55 | 130 | 0.423 | 7.7 | 13 | 1.69 |
| CAR58d | 4 hrs 50 min | 69 | 168 | 0.411 | 8.5 | 12.5 | 1.47 |
| CAR519f | 7 hrs | 61 | 142 | 0.430 | 7.9 | 13.1 | 1.66 |
| CAR511d | 7 hrs 01min | 73 | 178 | 0.410 | 8.4 | 13.2 | 1.57 |
| CAR521e | 10 hrs | 87 | 199 | 0.439 | 8.8 | 15.4 | 1.75 |
| CAR512d | 10 hrs 18 min | 106 | 246 | 0.431 | 9.6 | 14.0 | 1.46 |
| CAR512d (repeated) | 10 hrs 18 min | 104 | 263 | 0.395 | 9.9 | 14.0 | 1.41 |
| CAR55d | 15 hrs 01 min | 90 | 218 | 0.413 | 8.7 | 14.7 | 1.69 |
| CAR56e | 21 hrs 55 min | 107 | 252 | 0.425 | 9.6 | 14.0 | 1.46 |
| CAR57b | 32 hrs | 117 | 299 | 0.391 | 9.4 | 15.0 | 1.60 |

**Table 2:** Hysteresis cycle characteristics (measured at 400°C) of 13 distinct samples, each of them having been heated previously at 400 °C for a different duration (second column). The maximum applied field during hysteresis experiments was 0.190mT. The experimental procedure is described in §2.2.

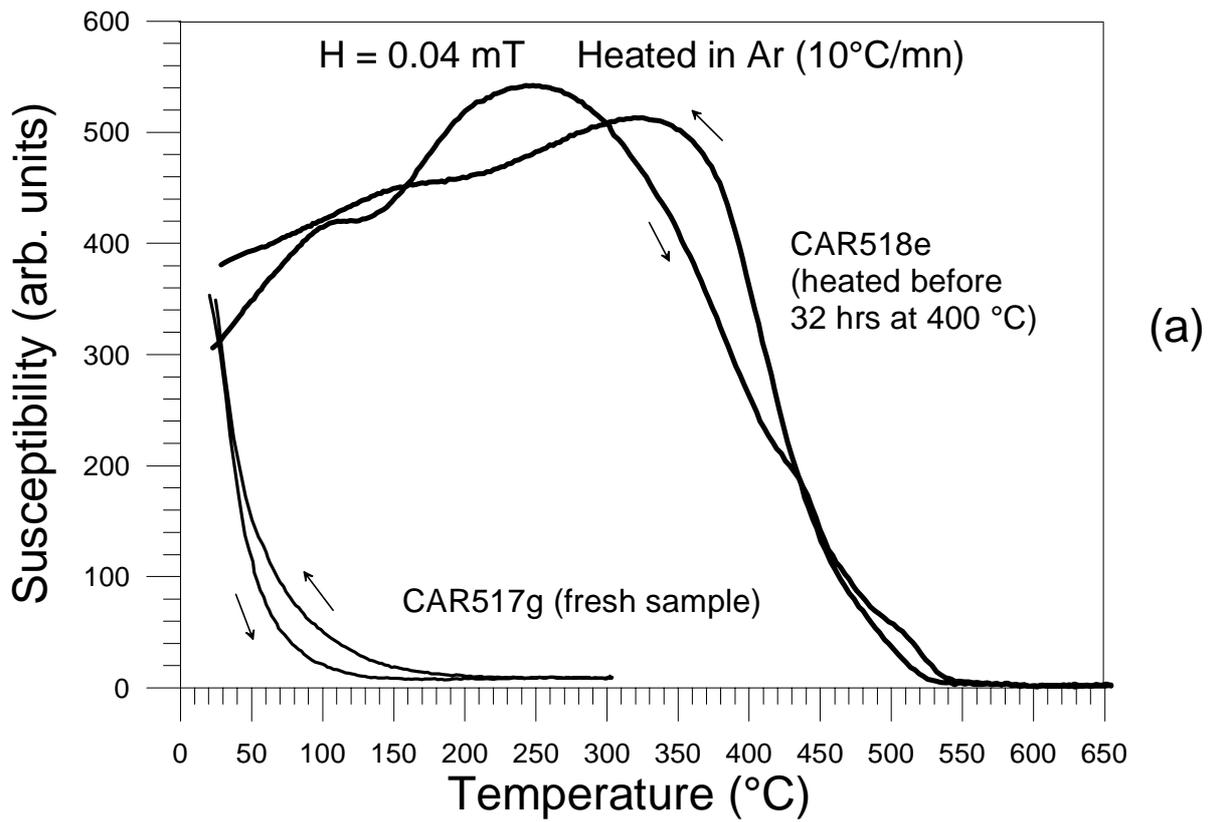

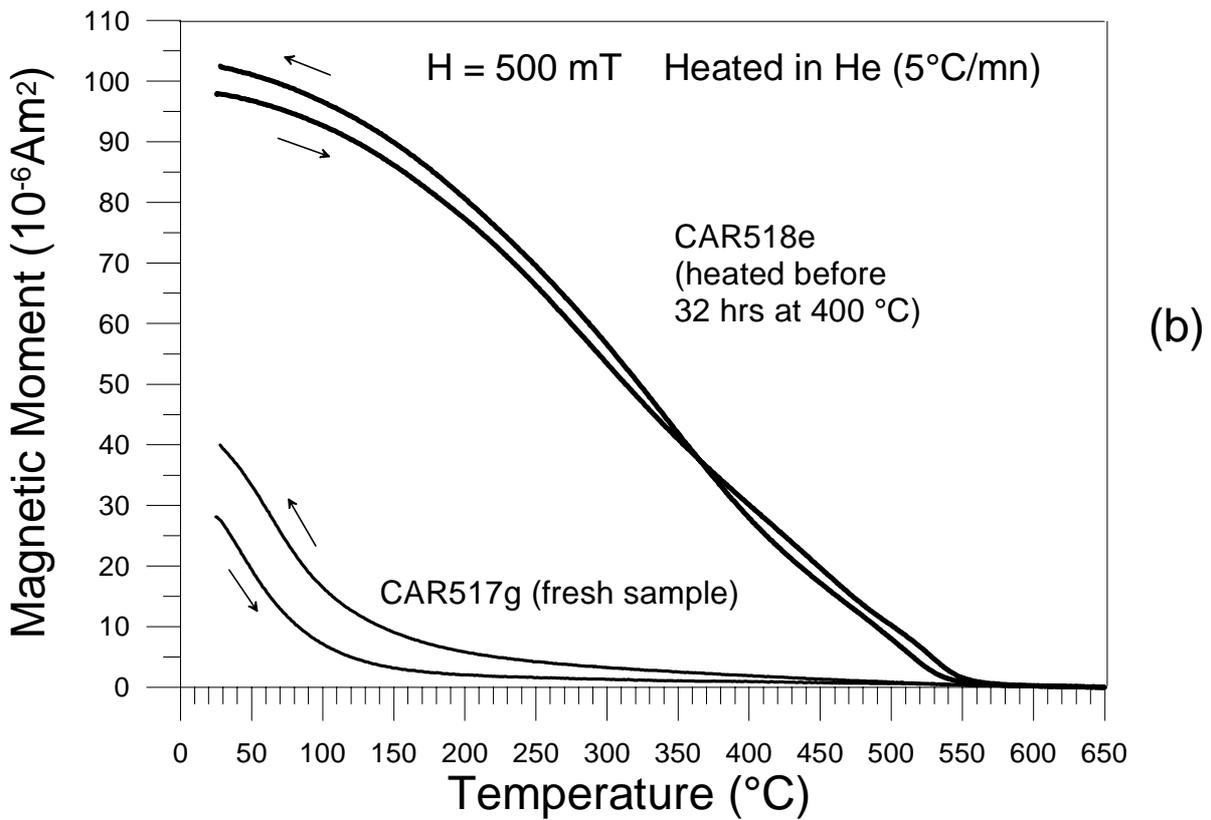

Figure 1. Low-field (a) and high-field (b) thermomagnetic curves of two distinct chips from sample CAR517g (previously unheated) and sample CAR518e (previously heated at 400°C for 32 hrs to acquire a CRM).

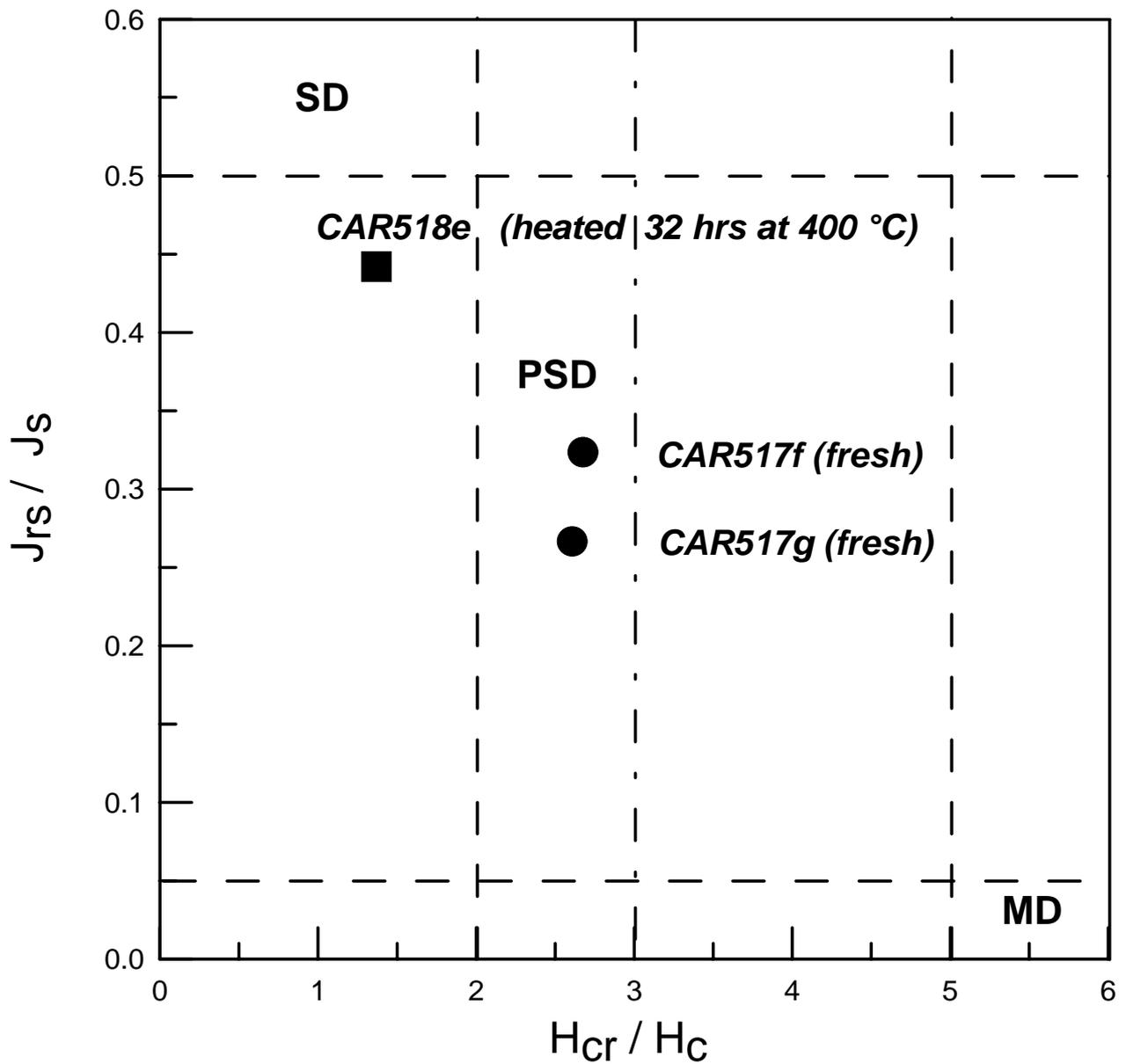

Figure 2. Day diagram (Jrs/Js versus Hcr/Hc) at room temperature for two fresh samples (CAR517f and g, full circles) and sample CAR518e (full square), the latter previously heated at 400°C for 32 hrs in order to acquire a CRM. Dashed lines indicates the approximate location of the boundaries of SD, PSD and MD domains for magnetite (Day, 1977; Dunlop, 2002). As the upper limit of the Hcr/Hc ratio for PSD grains varies with titanium content, this limit is indicated by a dash-and-dot line for composition TM60.

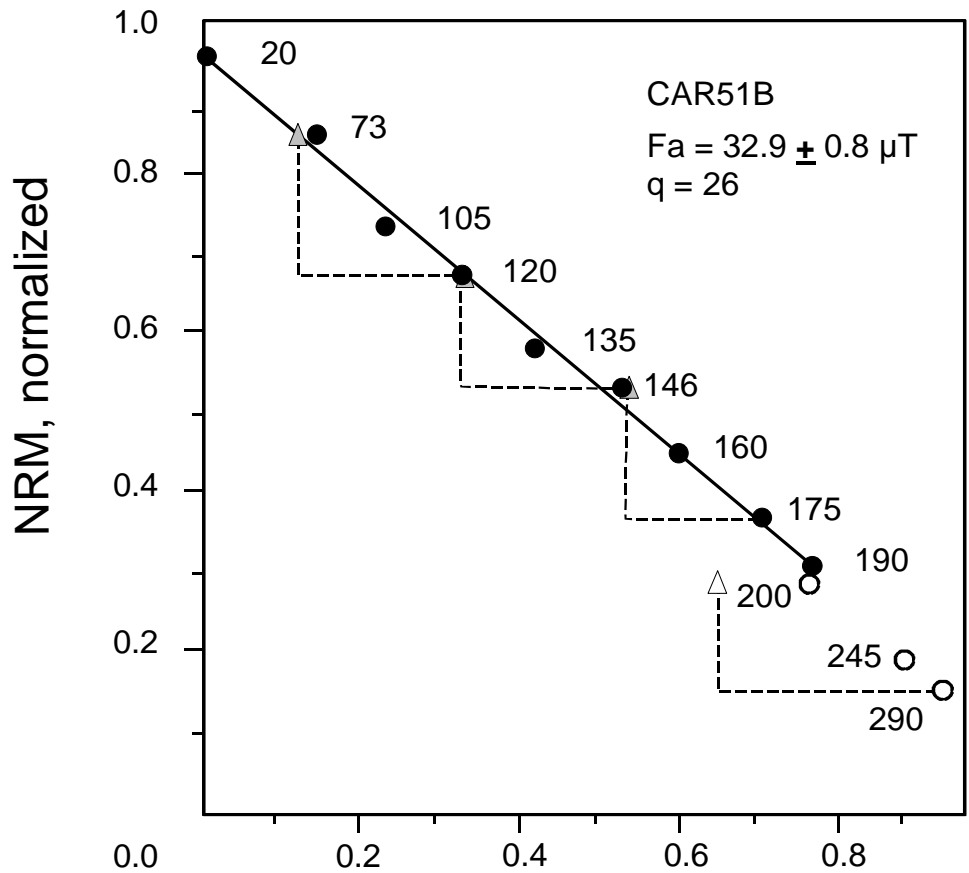

Figure 3. Example of NRM-TRM diagram for fresh material (sample CAR51B). Linearly extrapolated initial NRM is 5.40 10$^{-3}$Am$^2$/kg. Linearly extrapolated final TRM is 5.73 10$^{-3}$Am$^2$/kg. Laboratory field was 40µT. Fa is the estimated paleofield magnitude and q the quality factor (Coe et al., 1978).

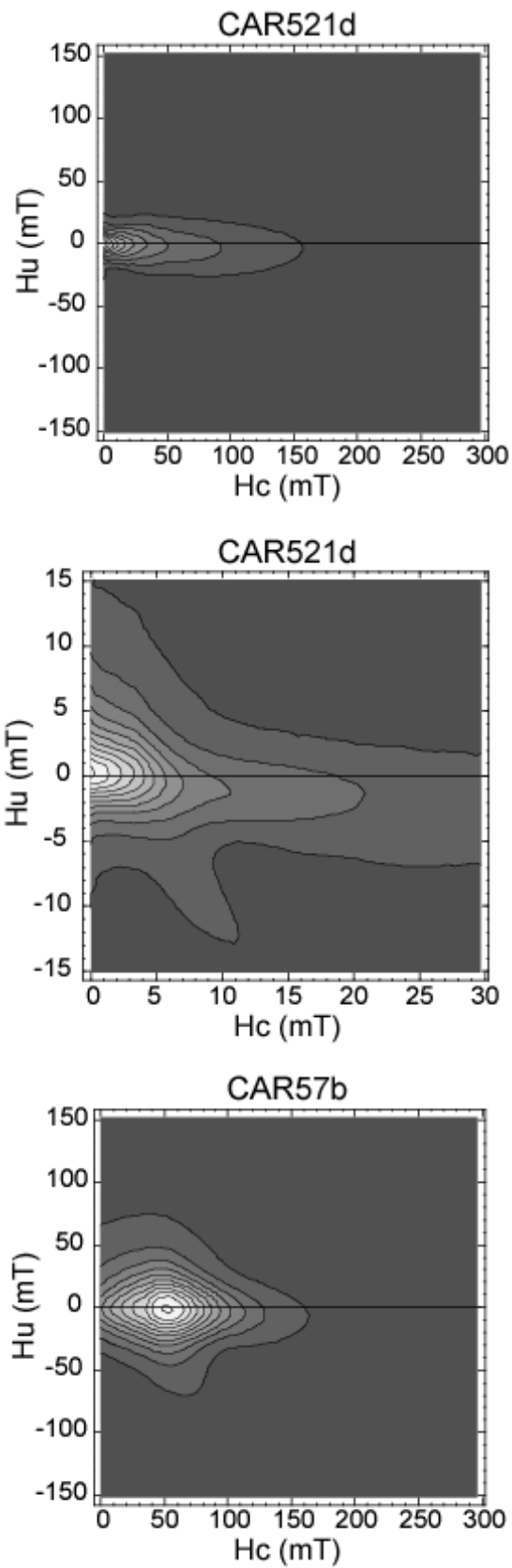

Figure 4. FORC diagrams for fresh sample CAR521d (upper two diagrams, the second corresponding to more detailed measurements) and sample CAR57b, the latter previously heated (first for 32 hours at 400°C for CRM acquisition, then for Thellier and hysteresis measurements at temperatures between 400°C and 580°C). The physical meaning of fields Hc and Hu are discussed in §3.1.

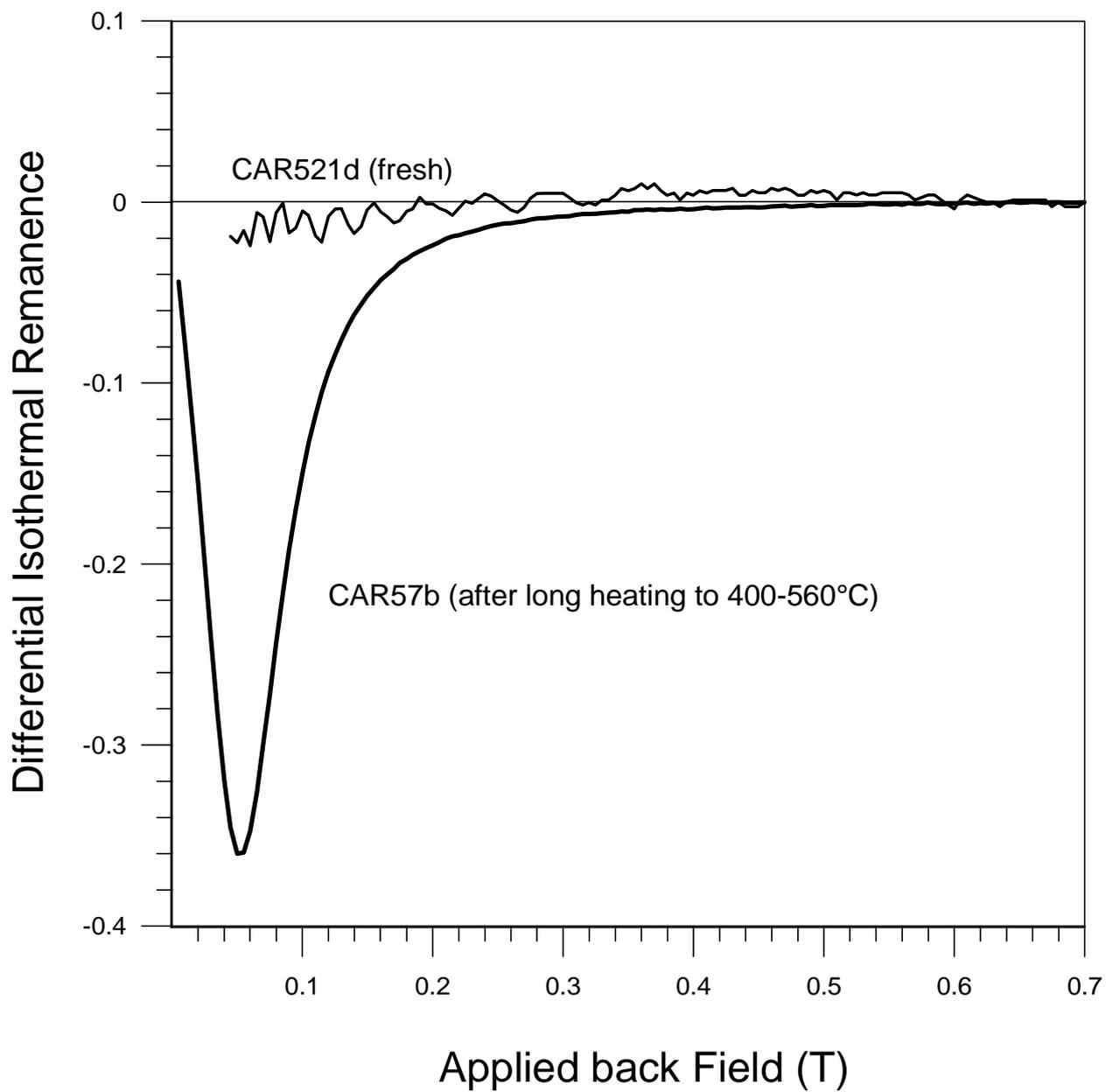

Figure 5. Differential IRM versus back dc field for chips of fresh sample CAR521d and previously heated sample CAR57b (see caption of Fig. 4 for description of previous thermal treatment).

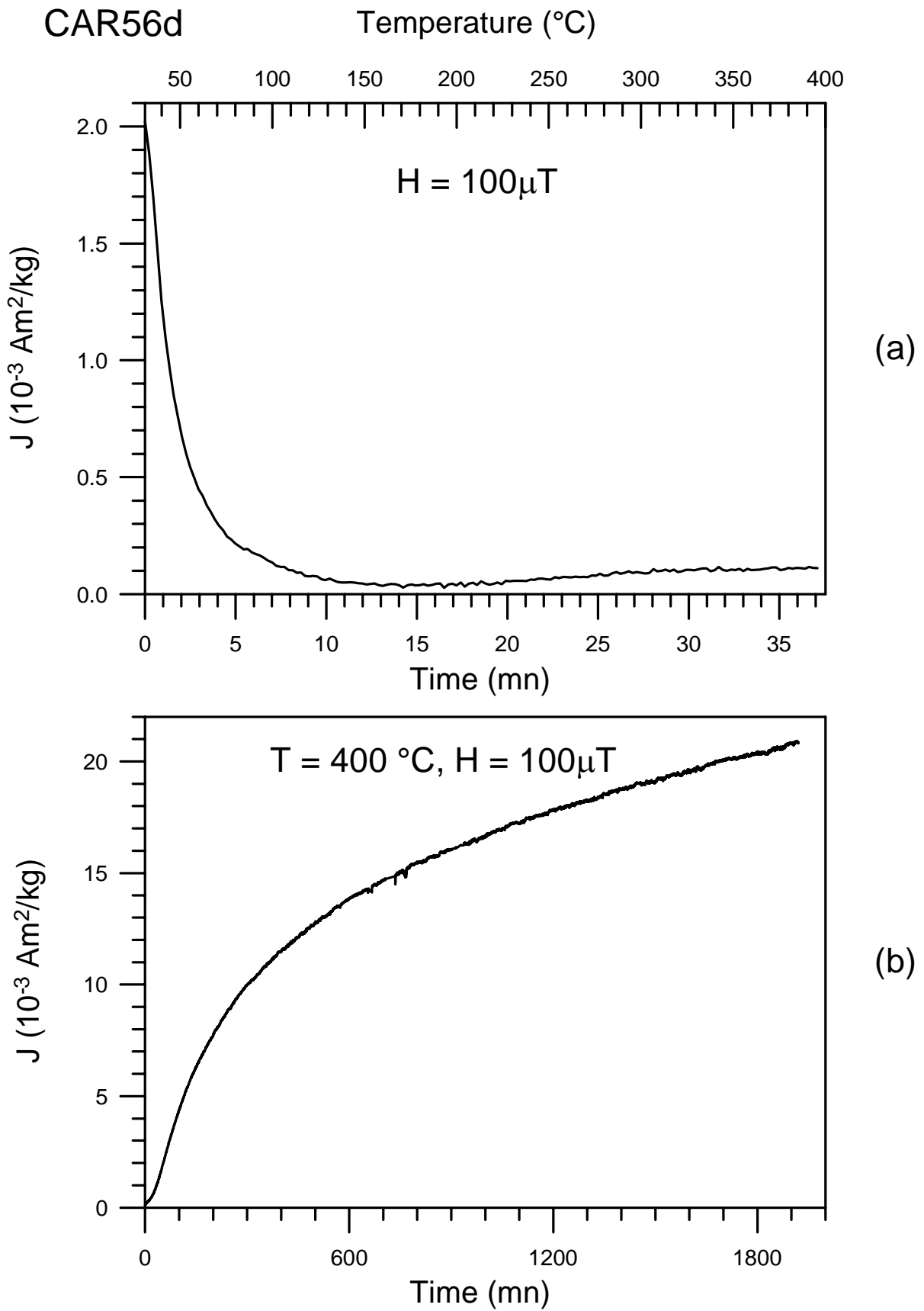

Figure 6. Thermal demagnetization curve of residual NRM (after AF demagnetization) (a), and curve of CRM acquisition versus time at 400°C (b), both being measured in field h=100μT, for sample CAR56d. In (b) the origin of the time axis is the beginning of the constant temperature stage.

CAR56d

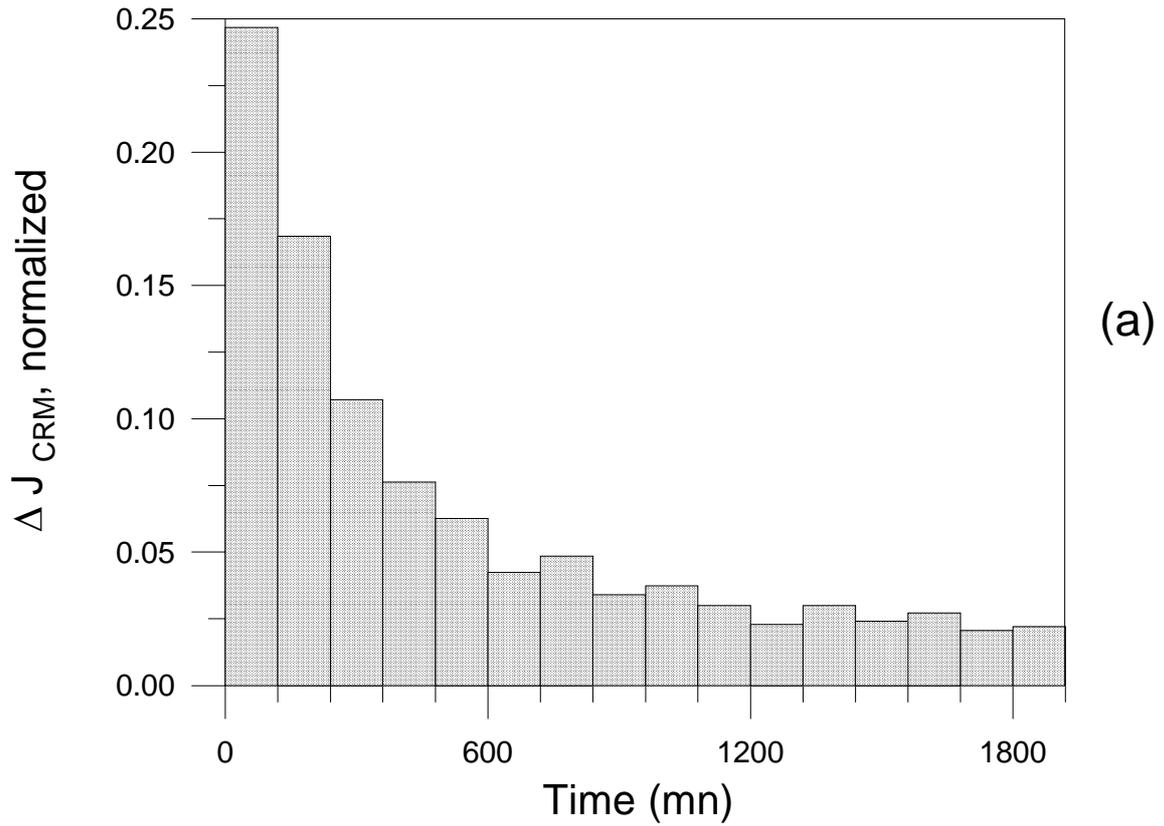

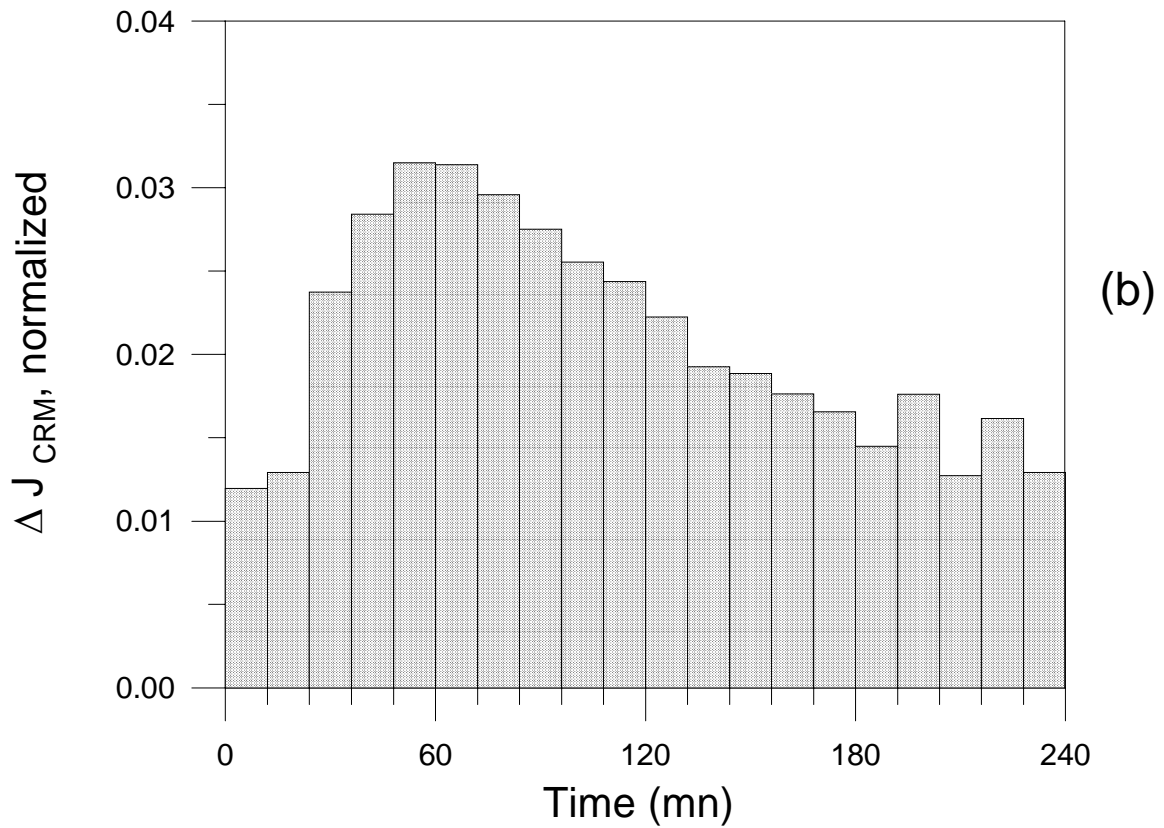

Figure 7. Histograms of partial CRM acquisition for sample CAR56d with time zero set at the beginning of the constant temperature stage. (a) covers the entire CRM acquisition stage while (b) is a zoom of the beginning of acquisition process. In both diagrams, ΔJcrm is normalized to the total CRM.

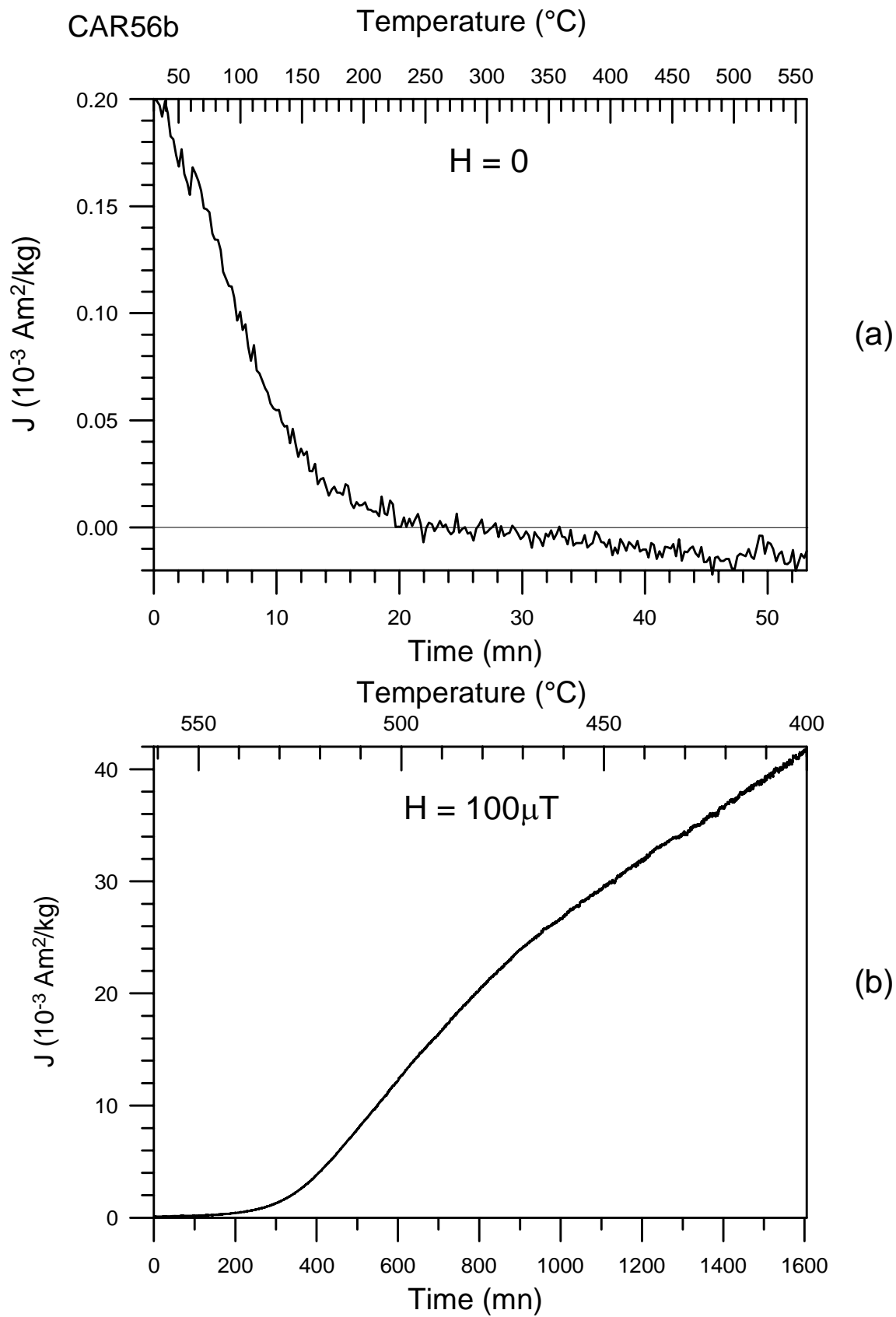

Figure 8. Demagnetization curve (a) of residual NRM (after AF demagnetization) and TCRM acquisition (b) for sample CAR56b. In diagram (b) time zero is set at the beginning of the steadily decreasing temperature stage. The progressive shift of the baseline visible in diagram (a) is due to the high sensitivity required for measuring the very small residual NRM.

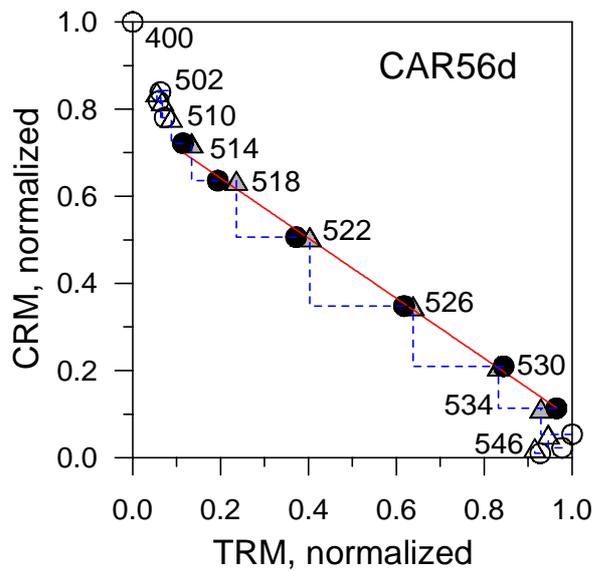

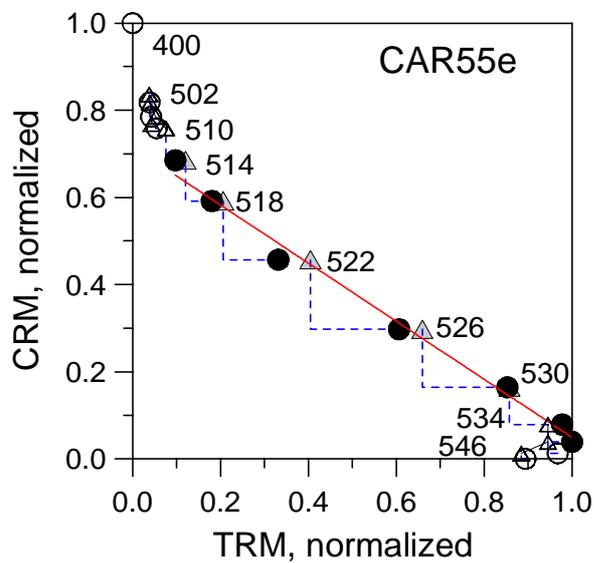

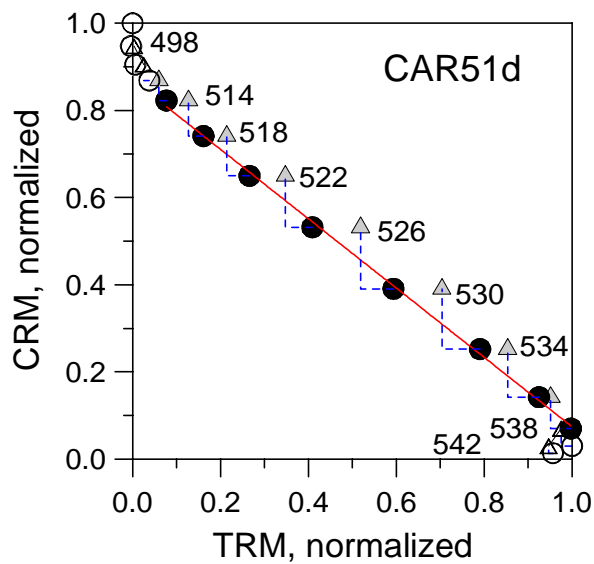

Figure 9. Examples of CRM-TRM diagrams (see §2.3 for description of experimental procedure). Full (empty) circles correspond to data points used (non used) for calculating the apparent "paleofield". Triangles indicate pTRM checks (Thellier and Thellier, 1959).

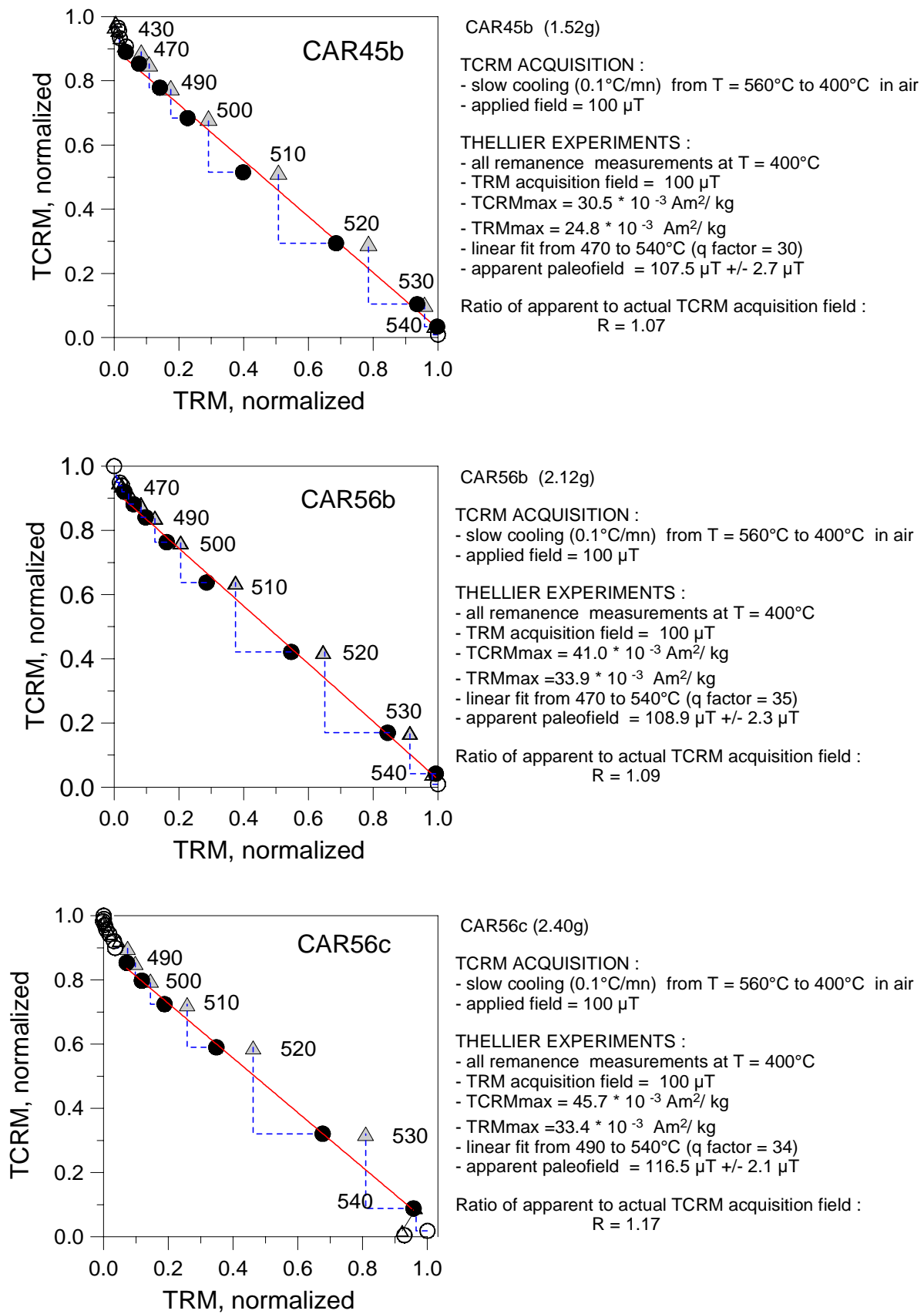

Fig. 10. Examples of TCRM-TRM diagrams (see §2.3 for description of experimental procedure). Full (empty) circles correspond to data points used (non used) for calculating the apparent "paleofield". Triangles indicate pTRM checks (Thellier and Thellier, 1959).

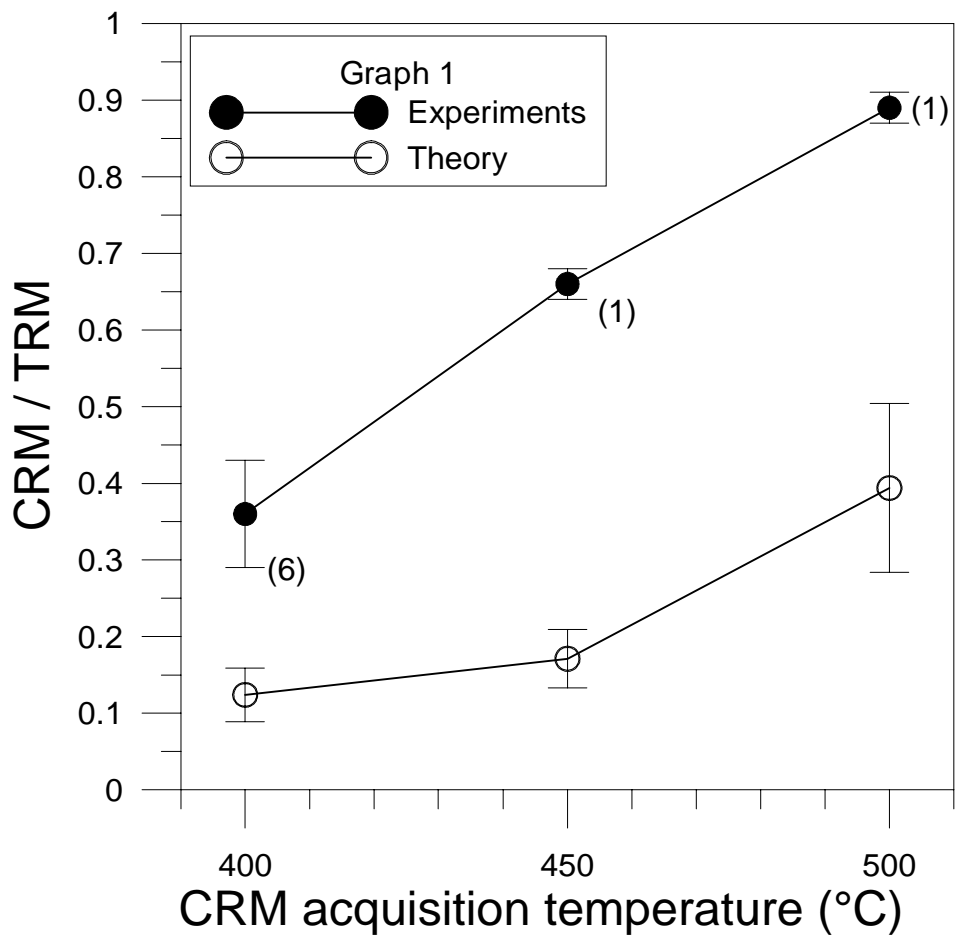

Figure 11. Mean CRM/TRM ratio versus the temperature of CRM acquisition. The experimental data (full circles) are from Table 1. Labels by each circle indicate the number of experimental data. Error bars correspond either to the apparent "paleointensity" error (at 450 and 500°C) or the standard deviation of apparent "paleointensity" determinations (at 400°C). Theoretical ratio and associated error bar were calculated from equation (11) assuming a TRM blocking temperature of 525±5°C.

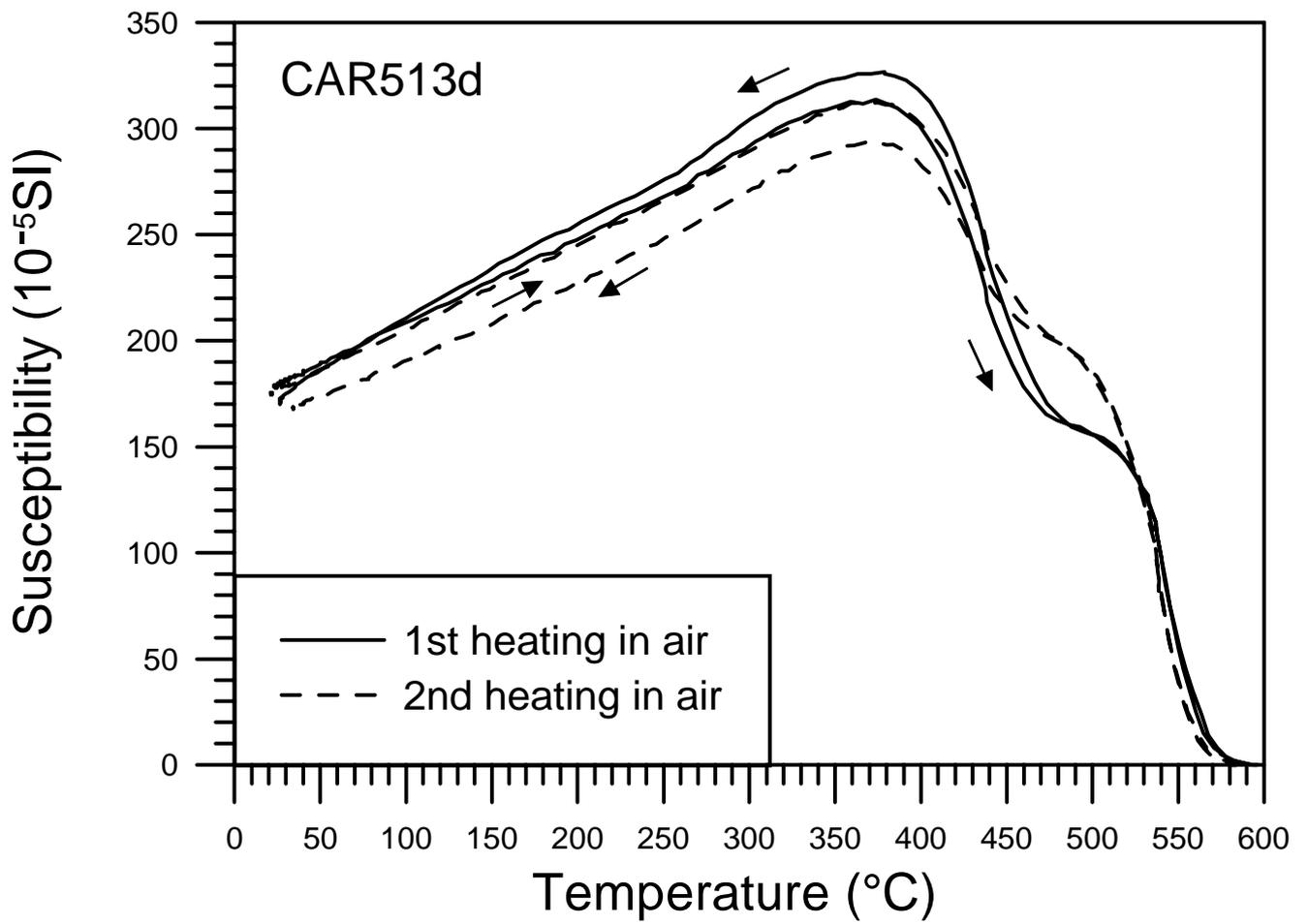

Figure 12. Two successive low-field thermomagnetic cycles in air for sample CAR513d. Previously, this sample had been heated at 400°C for 32 hrs to acquire a CRM and then subjected to Thellier experiments.

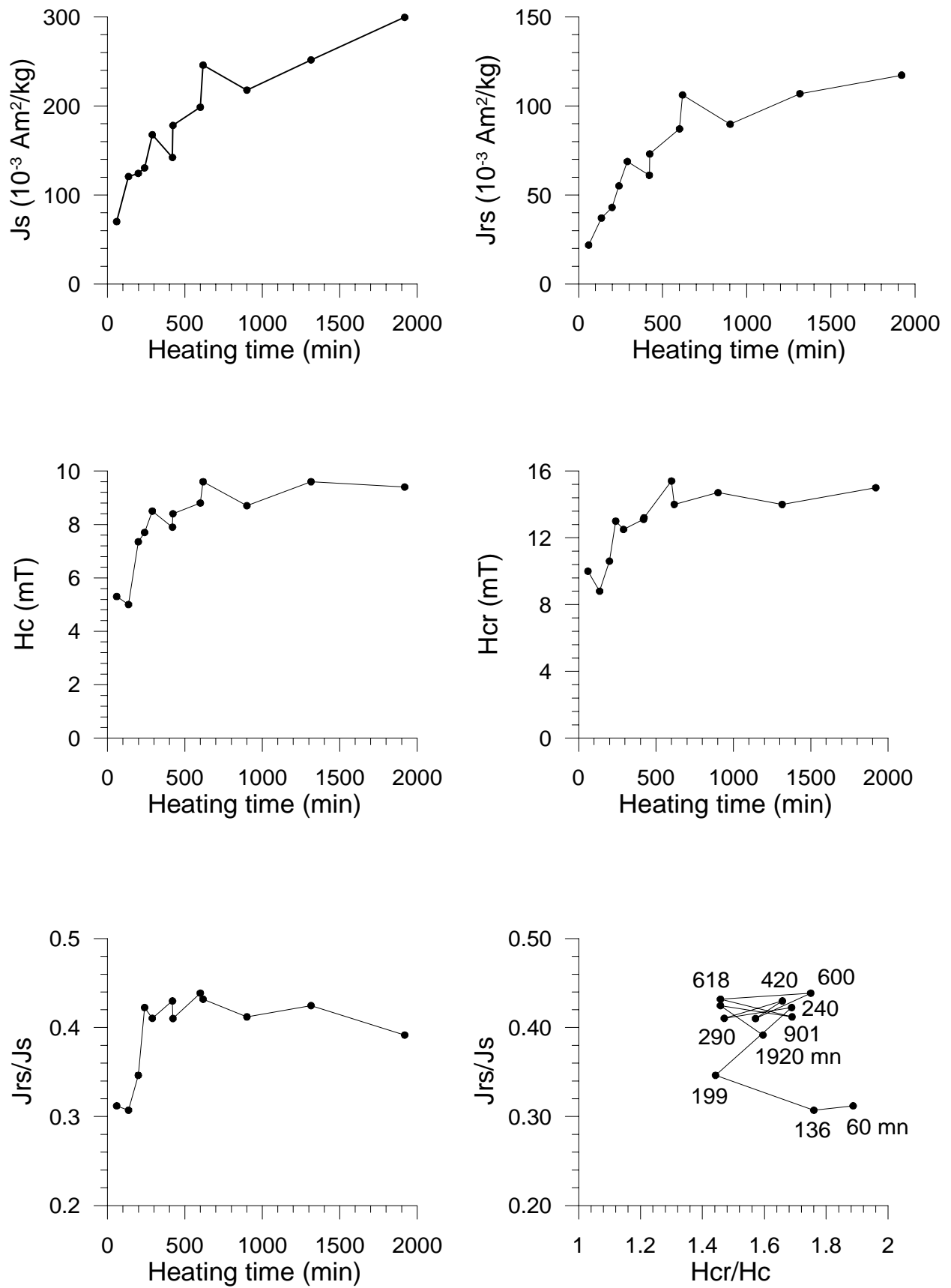

Figure 13. Changes in the main characteristic quantities of hysteresis cycle at 400°C versus the duration of heating time at this temperature, as measured on 13 distinct samples from hand sample CAR5. The relative magnetic homogeneity of that set of samples, attested by the regularity of the observed changes with heating time, results from the rigorous selection criteria of our samples (§ 2.1).

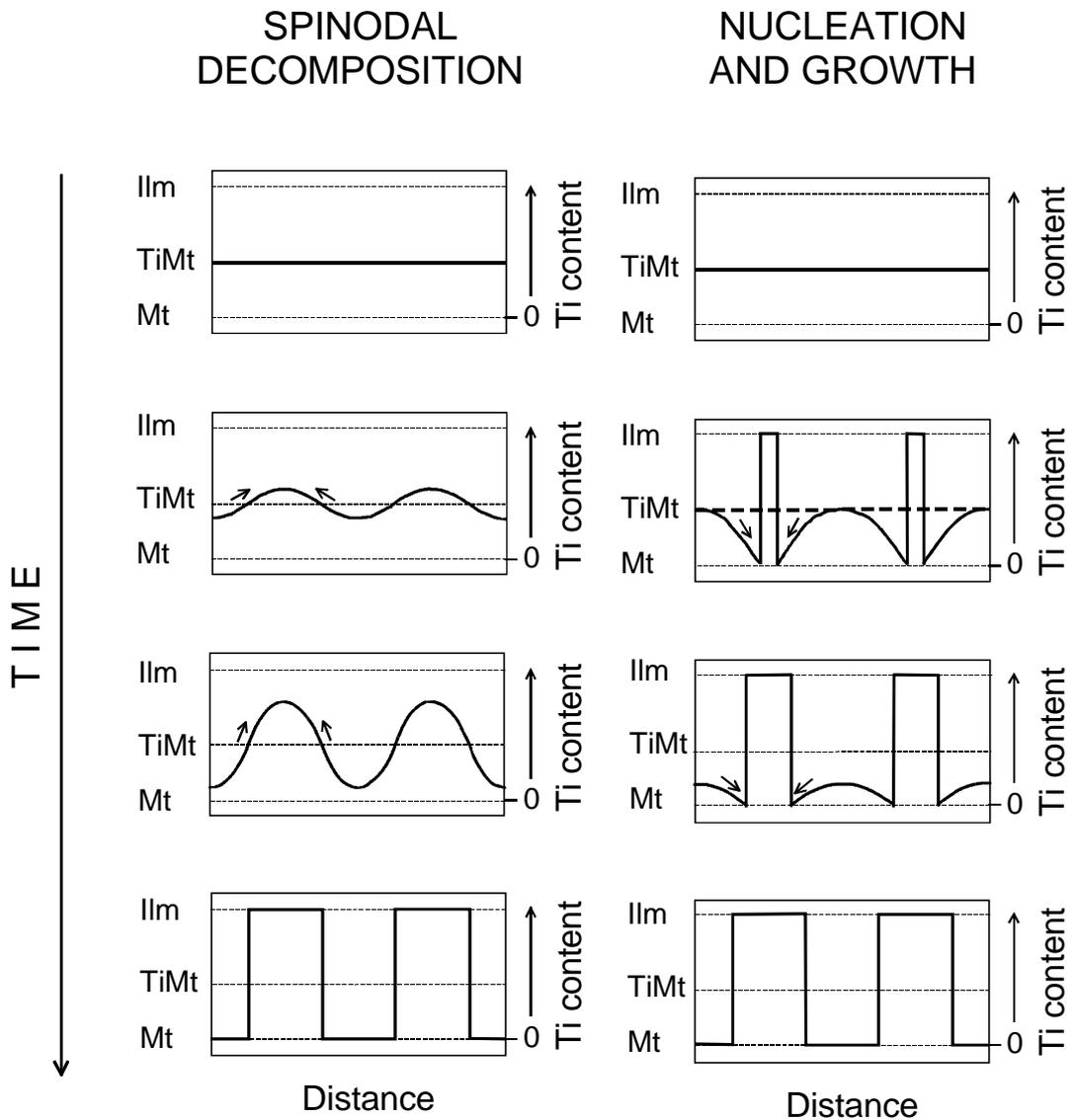

Figure 14. Oxyexsolution in titanomagnetite: a comparison of spinodal decomposition and nucleation and growth (modified from Putnis, 1992).

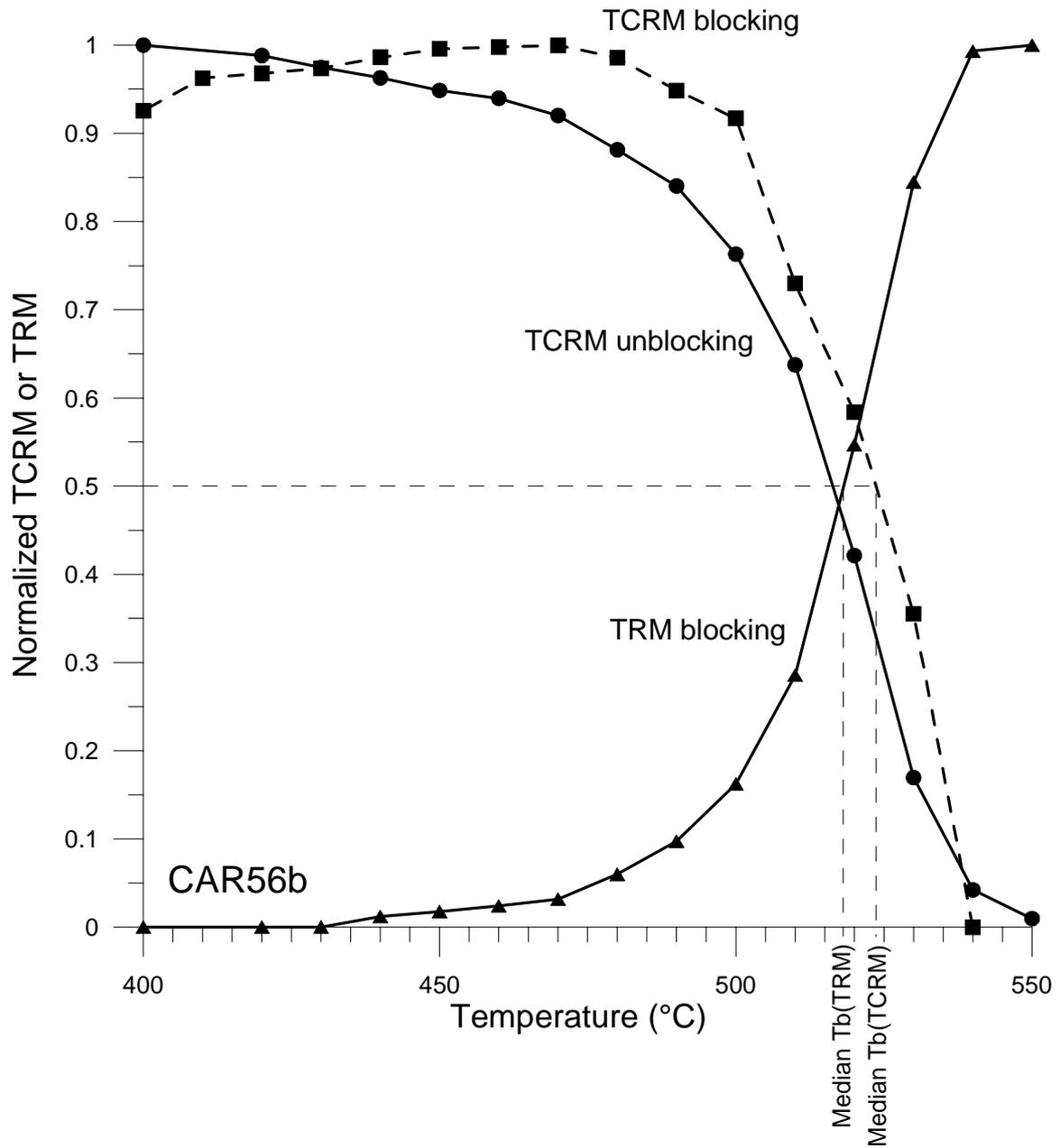

Figure 15. TCRM blocking (dashed line, calculated), TCRM unblocking (solid line, measured), and TRM blocking (solid line, measured) for sample CAR56b.

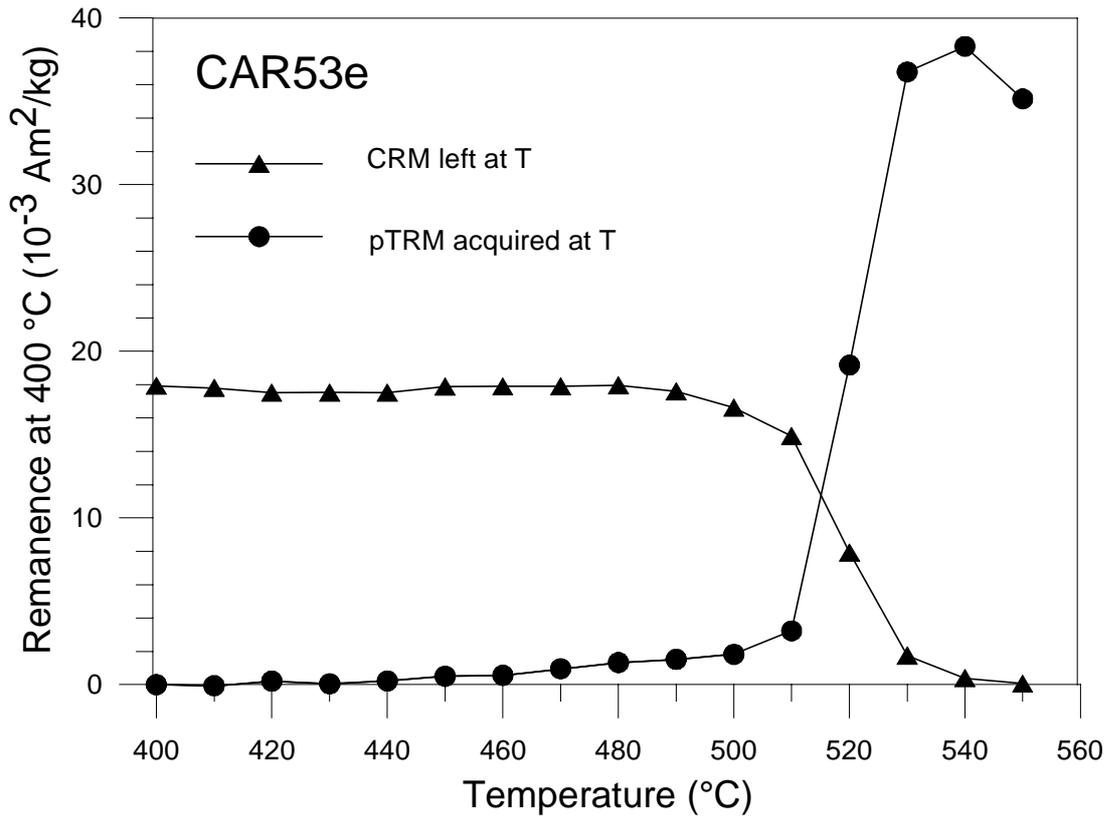
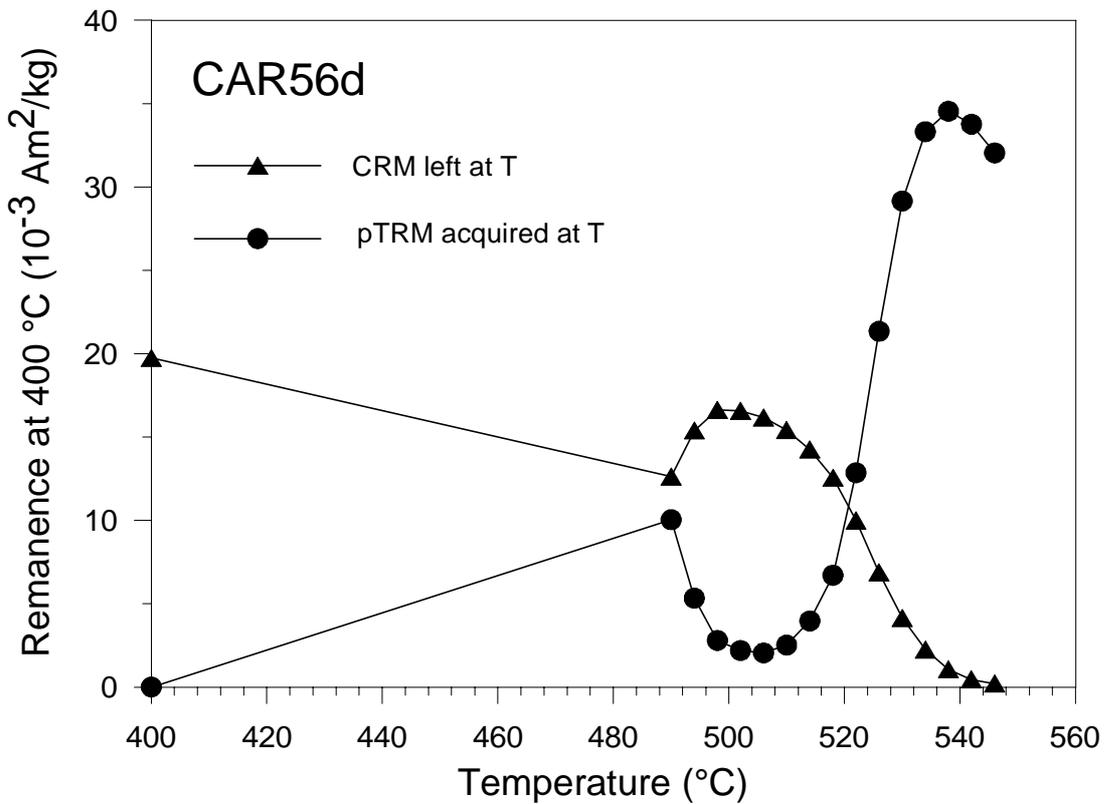

Figure 16. CRM and TRM changes versus temperature during Thellier experiments in case of (a) 10°C step heating from 400°C to 550°C or (b) a rapid heating (4 °C/min) to 492°C followed by 4°C step heating up to 546°C.

CAR518e

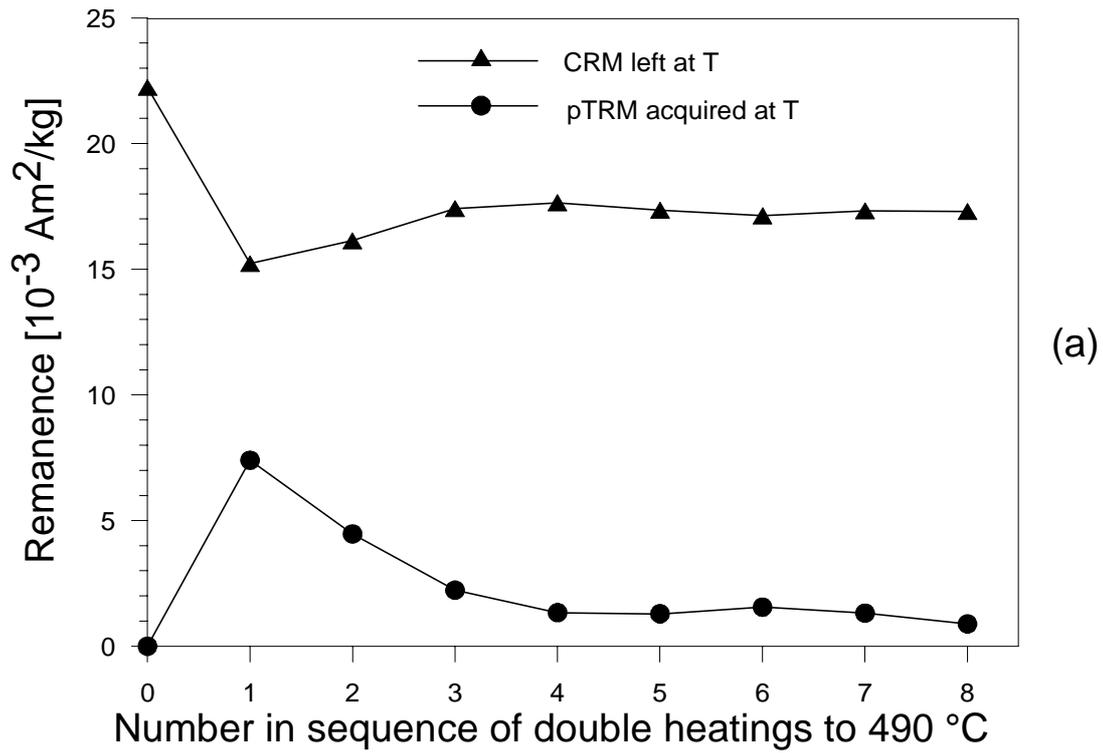

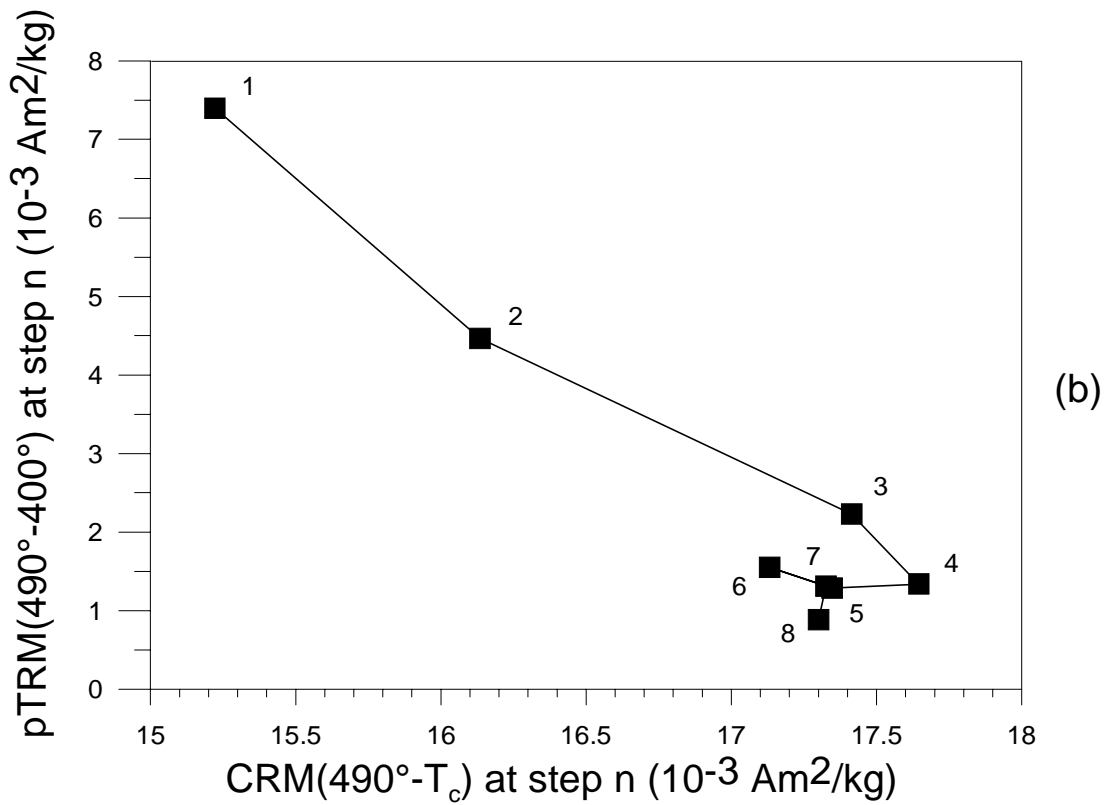

Figure 17. Effect of repetitive double heatings at 490°C on the CRM(490°-Tc) and the TRM (490°-400°) of sample CAR518c (a). Plot of TRM (490°-400°) versus CRM(490°-Tc) (b). The dot labels indicate the sequence number of successive double heatings.